\documentclass[]{aastex631}
\usepackage{subfigure}

\submitjournal{ApJ}
\accepted{April 2026}
\shorttitle{Kinematics of LIRGs at z = 0.5--0.6}
\shortauthors{Eleazer et al.}

\begin{document}
\author[0000-0002-4690-4502]{Eleazer, Miriam}
\affiliation{Department of Astronomy, University of Massachusetts, Amherst, MA 01003, USA}

\author[0000-0001-8592-2706]{Pope, Alexandra}
\affiliation{Department of Astronomy, University of Massachusetts, Amherst, MA 01003, USA}

\author[0000-0002-1917-1200]{Sajina, Anna}
\affiliation{Department of Physics \& Astronomy, Tufts University, Medford, MA 02155, USA}

\author[0000-0003-1710-9339]{Yan, Lin}
\affil{Caltech Optical Observatories, California Institute of Technology, Pasadena, CA 91125, USA}

\author[0000-0002-5830-9233]{Young, Jason}
\affiliation{SETI Institute, Mountain View, CA 94043, USA}
\affiliation{Department of Astronomy, University of Massachusetts, Amherst, MA 01003, USA}

\author[0000-0002-8909-8782]{Alberts, Stacey}
\affiliation{Space Telescope Science Institute, Johns Hopkins University, Baltimore, MD 21218, USA}
\affiliation{Department of Astronomy, University of Arizona, Tucson, AZ 85721, USA}

\author[0000-0003-3498-2973]{Armus, Lee}
\affiliation{Department of Astronomy, California Institute of Technology, Pasadena, CA 91125, USA}

\author[0000-0002-0729-2988]{Coppin, Kristen}
\affiliation{Department of Physics, Astronomy \& Mathematics, University of Hertfordshire, Hatfield, AL10 9EU, UK}
\affiliation{Centre for Astrophysics Research, Astronomy \& Mathematics, University of Hertfordshire, Hatfield, AL10 9EU, UK}

\author[0000-0002-5782-9093]{Dale, Daniel}
\affiliation{Department of Physics \& Astronomy, University of Wyoming, Laramie, WY 82071, USA}

\author[0000-0003-1748-2010]{Farrah, Duncan}
\affiliation{Department of Physics \& Astronomy, University of Hawai'i at Manoa, Honolulu, HI 96822, USA}
\affiliation{Institute for Astronomy, University of Hawai'i at Manoa, Honolulu, HI 96822, USA}

\author[0000-0003-2374-366X]{Gon\c calves, Thiago}
\affiliation{Department of Astronomy, Universidade Federal do Rio de Janeiro, Rio de Janeiro, RJ, 21941, BR}

\author[0000-0002-6149-8178]{McKinney, Jed}
\affiliation{Department of Astronomy, University of Texas at Austin, Austin, TX 78712, USA}

\author[0000-0001-5783-6544]{Nesvadba, Nicole}
\affiliation{Department of Astronomy, Universit\'e C\^ote d'Azur, Nice, 06000, FR}

\author[0000-0002-3471-981X]{Ogle, Patrick}
\affiliation{Space Telescope Science Institute, Johns Hopkins University, Baltimore, MD 21218, USA}

\author[0000-0001-8245-7669]{Popescu, Roxana}
\affiliation{Department of Astronomy, University of Massachusetts, Amherst, MA 01003, USA}

\author[0000-0002-3158-6820]{Veilleux, Sylvain}
\affiliation{Department of Astronomy, University of Maryland, College Park, MD 20742, USA}
\affiliation{Joint Space-Science Institute, NASA Goddard Space Flight Center, Greenbelt, MD 20771, USA}
\title{Halfway to the Peak: Kinematic Signatures of Stable Rotating Disks in Luminous Infrared Galaxies at z=0.5-0.6}

\begin{abstract}
We present a kinematic study of six infrared-luminous galaxies observed with the Mid-InfraRed Instrument Medium-Resolution Spectrometer (MIRI/MRS) onboard JWST. These galaxies lie at $z = 0.5$--$0.6$, midway between the present day and the peak of cosmic star formation. Our sample spans a range of star formation (SF) and active galactic nucleus (AGN) contributions to the mid-infrared emission. We characterize the dynamical state of these IR-luminous galaxies and assess how AGN activity influences the kinematics of the interstellar medium. Using mid-IR atomic lines, we map galaxy kinematics beyond the local Universe for the first time. The spatial resolution of MIRI/MRS (3.0 kpc for 0.46$\arcsec$ at z $\sim$ 0.55) allows us to resolve the internal kinematics of our targets. We compute kinematic maps in three different emission lines ([Ar II]6.99$\mu$m, [Ne II]12.81$\mu$m, and H$_2$ 0-0 S(5)6.91$\mu$m). Using the [Ar II]6.99$\mu$m kinematic maps, we derive rotation curves for these sources. All galaxies exhibit ordered rotation, with \(V/\sigma \geq 2\), consistent with stable disks. Although some show minor disturbances, we find no strong evidence for recent major mergers or galaxy-wide ionized outflows. We find no correlation between \(V/\sigma\) and AGN fraction, suggesting AGN activity does not significantly disrupt global kinematics or that disk disruption is not required to trigger AGN. However, galaxies with higher AGN fractions show elevated central dispersions, indicating localized turbulence, possibly due to AGN feedback, stellar feedback, accretion or bulge structure. These IR-luminous galaxies likely represent mature, rotationally supported disks, with AGN activation occurring after disk assembly.
\end{abstract}

\keywords{Luminous infrared galaxies --- JWST --- Galaxy kinematics --- Galaxy rotation --- Galaxy evolution --- Active galactic nuclei} 

\section{Introduction} \label{sec:intro}

Luminous Infrared Galaxies (LIRGs, $L_{IR} \geq 10^{11} L_{\odot}$, star formation rates, SFR $\gtrsim 10M_{\odot}/$yr) and Ultraluminous Infrared Galaxies (ULIRGs, $L_{IR} \geq 10^{12} L_{\odot}$, star formation rates, SFR $\gtrsim 10^{2}M_{\odot}/$yr) and have spectral energy distributions (SEDs) that are dominated by thermal dust emission, caused by the reprocessing of UV photons emitted by young, massive stars and/or active galactic nucleus (AGN) heating (Sanders \& Mirabel 1996; Kartaltepe et al.~2010; Alonso-Herrero et al.~2012; U et al.~2012; Bellocchi et al.~2013, Pérez-Torres et al.~2020). While relatively rare in the local universe (Sanders \& Mirabel 1996; Wang et al.~2006; Kartaltepe et al.~2010; Alonso-Herrero et al.~2012; Pérez-Torres et al.~2020), LIRGs begin to dominate the IR background at a redshift of $\sim$ 1, making them major contributors to the cosmic star formation rate density (SFRD) of the universe (e.g.~Le Floc’h et al.~2005; Murphy et al.~2011; Magnelli et al.~2013).

IR-luminous galaxies track several different key stages of galaxy evolution and display diverse morphologies across cosmic time. At z $\sim 0$, a significant fraction of IR-luminous galaxies are associated with mergers or interactions, which are also more likely to host an AGN than a normal galaxy (Sanders \& Mirabel 1996). This is consistent with models that mergers are triggering AGN activity (Hopkins et al.~2008), although a portion ($\sim$35\%) appear to be isolated disk galaxies (Stierwalt et al.~2014). This suggests that even in the local universe, major mergers are not a prerequisite for LIRG-level star formation or AGN activity. At z $\sim$ 0.55, where gas fractions and star formation rates are higher, we might expect these secularly evolving, rotationally supported IR-luminous galaxies to be even more prevalent (Crespo G\'omez et al.~2021). 

In this context, spatially resolved kinematic studies are particularly powerful. A 2D kinematic characterization of galaxies can probe the physical processes that shape their formation and evolution. It provides a valuable diagnostic framework to infer the primary source of dynamical support (Puech et al.~2007; F\"orster-Schreiber et al.~2009; Bellocchi et al.~2013, 2018; Glazebrook 2013; Crespo Gómez et al.~2021), differentiate between relaxed, virialized systems and merger-driven dynamics (Flores et al.~2006; Shapiro et al.~2008; Bellocchi et al.~2012, 2013, 2016), and identify and characterize radial motions linked to feedback mechanisms such as outflows, where gas is expelled from the galaxy due to energetic processes like intense star formation or AGN activity, impacting the galaxy's evolution  (Shapiro et al.~2009; Rupke \& Veilleux 2013; Bellocchi et al.~2013; Arribas et al.~2014; Fluetsch et al.~2019). In LIRGs and ULIRGs, such outflows are often observed alongside disturbed, non-rotational kinematics, reflecting the complex interplay between merger-driven dynamics and feedback from both star formation and AGN activity (Veilleux et al.~2005; Veilleux et al.~2020).

At higher redshifts ($z =$ 2--3), galaxy disks have been found to be more turbulent, with higher velocity dispersion, and less rotationally supported than disks at z $<$ 1 (Wisnioski et al. 2015, 2025; Rizzo et al. 2024). From $z \approx 1$ to $z \approx 0$, observations consistently show that the intrinsic gas velocity dispersion ($\sigma_{gas}$) decreases linearly with cosmic time (\"Ubler et al. 2019; Mai et al. 2024). This decreasing global average $\sigma_{gas}$ is consistent with the simultaneous decrease in specific star formation rate (sSFR) and gas fraction over time (Kassin et al. 2012; Glazebrook et al. 2013; Wisnioski et al. 2015, 2025; \"Ubler et al. 2019; Mai et al. 2024). This indicates that, while high-redshift galaxy disks are rotationally supported on average (see the right panel of Figure \ref{fig:vsig_AGN}, which shows the median $z \sim 2$ star-forming disks from F\"orster-Schreiber et al.~2018), they are dynamically less stable and more turbulent than their $z=0$ counterparts, which are generally more settled after consuming much of their gas content.

The Mid-InfraRed Instrument/Medium-Resolution Spectrometer (MIRI/MRS) on JWST can map galaxy kinematics using mid-IR tracers that are less affected by dust attenuation. Galaxies at z $\sim$ 0.5 have angular sizes that fall completely within the field-of-view of MRS to enabled detailed analysis of kinematic structures at a transitional redshift. This redshift range lies halfway, in cosmic time, to the peak epoch of cosmic star formation (z $\sim$ 2), and is thought to represent a transitional phase in galaxy evolution. At $z =$ 0.5--0.6, the cosmic star formation rate density was approximately 3 times higher than it is today (Madau \& Dickinson, 2014). Observationally, a significant fraction of quenching in field galaxies appears to occur around z $\sim$ 0.5 (Pandya et al.~2017). From a theoretical perspective, cosmological simulations and semi-analytic models (Hopkins et al.~2006) suggest that the frequency of major mergers declines significantly after z $\sim$ 1, meaning that many massive galaxies have already undergone their most disruptive merger events by z $\sim$ 0.5. Thus, determining whether IR-luminous galaxies at z $\sim$ 0.5 are dynamically settled or still shaped by interactions provides a key opportunity to investigate the mechanisms driving star formation suppression and the transition to quiescence.

In this paper, we focus on five LIRGs and one ULIRG at $z =$ 0.5--0.6 that span a wide range of mid-infrared (MIR) AGN fractions. \footnote{The (MIR) AGN fraction is defined as the proportion of the total MIR emission that is contributed by the active galactic nucleus.}
By selecting galaxies with a range of MIR AGN fractions, we can explore how AGN activity may influence key galaxy properties such as morphology and ionized gas kinematics. Specifically, we analyze the kinematics of these systems to determine the dominant sources of dynamical support; whether these galaxies are rotationally supported, dispersion dominated, or exhibit signs of major mergers or interactions.

This paper is organized as follows. Section 2 presents the sample and observational details, including the data collection and reduction methods used for the JWST MIRI/MRS observations. In Section 3, we outline the analysis techniques used for spectral extractions, emission line modeling, and kinematic map-making. Section 4 presents the results, including kinematic maps, velocity and dispersion curves, $V/\sigma$ measurements, and relative dispersion fraction (see \S 4.2.3 for details) with a focus on the impact of AGN activity on the dynamics of IR-luminous galaxies. In Section 5, we dive into the discussion, interpreting our findings in the context of galaxy evolution and the interplay between star formation and AGN in these galaxies. Finally, in Section 6, we conclude with a summary of our key insights. In this work, a cosmology of $\Omega_m = 0.3$, $\Omega_\Lambda = 0.7$, and $H_0 = 70$ $\rm{km}$ s$^{-1}$ $\rm{Mpc}^{-1}$ is assumed.

\section{Sample \& Observations} \label{sec:sample}

Our JWST MIRI sample comes from the Halfway to the Peak Program (GO 1762, Young et al.~2023; Sajkov et al.~2025; Sajina et al.~2025) consisting of five LIRGs and one ULIRG at redshift $z =$ 0.5--0.6, spanning a range from star formation to AGN-dominated systems based on existing Spitzer/InfraRed Spectrograph (IRS) spectroscopy. The mid-IR spectral regime is a particularly good probe of what is producing the IR luminosity since it has contributions from both star formation and AGN activity, each with a unique signature at these wavelengths. Our sample is a subset of galaxies selected at 24 $\mu$m in the First Look Survey (FLS) from Kirkpatrick et al.~(2015), and one galaxy from a similar dataset in COSMOS (Fu et al.~2010). 


The mid-infrared (MIR) AGN fractions, used for sample selection based on the IRS data, are published (Kirkpatrick et al.~2015; Fu et al.~2010), and are calculated following the methodology outlined in Pope et al.~(2008). In this approach, the MIR spectrum is decomposed using a simple spectral energy distribution (SED) model comprising three primary components: polycyclic aromatic hydrocarbon (PAH) emission template, an AGN power-law continuum, and extinction characterized by prominent silicate absorption features. Once these components are separated, the MIR AGN fraction is determined by integrating the AGN component and comparing it to the total mid-infrared luminosity. An example of this SED model decomposition is shown in Figure~13 of Kirkpatrick et al.~(2015). We verified that the MIR AGN fractions derived from the old Spitzer/IRS spectra (Kirkpatrick et al.~2015) and our new JWST/MRS spectra are consistent (Young et al.~in prep). 
Our targets are listed in Table 1 including the coordinates, redshift, MIR AGN fraction and total IR (8--1000$\,\mu$m) luminosity. We also indicate the HST imaging band available for each source and the morphological classification from the literature. Our sample consists of two SF-dominated sources (COSMOS 1, FLS 1) while the remaining have significant AGN contribution.

\begin{table}[h]
    \centering
    \begin{minipage}[t]{\textwidth}
    \centering
    \begin{tabular}{cccccccc}
         ID&  R.A.&  Dec&  &  log(L$_{IR}$)&  MIR AGN & HST & Morphological\\
         & (h m s) & ($^o$ ' $\arcsec$) & z & (L$_\odot$) &fraction (\%) & Band & Classification\\
         \hline 
         COSMOS 1&  10:01:09.54&  +02:09:36.00&  0.6136 $\pm$ 5E-4&  11.92&  15  &  ACS WFC F814W & n/a\\
         FLS 1&  17:24:58.33&  +58:55:59.63&  0.5377 $\pm$ 5E-4&  12.02&  17  & NICMOS WFC3 FW160 & merger, early\\
         FLS 2&  17:24:58.33&  +59:57:03.06&  0.4943 $\pm$ 5E-4&  11.61&  56  & ACS WFC F475W & merger, late\\
         FLS 4&  17:23:01.44&  +59:58:43.61&  0.5234 $\pm$ 5E-4&  11.74&  67  & NICMOS WFC3 FW160 & spiral\\
         FLS 3&  17:21:18.31&  +58:46:01.56&  0.5553 $\pm$ 5E-4&  11.83&  91  & n/a & n/a \\
         FLS 6&  17:13:24.16&  +58:55:43.28&  0.6098 $\pm$ 5E-4&  11.96&  91  & ACS WFC F475W & merger, late\\
         
    \end{tabular}
    \caption{Properties of our sample, ordered by increasing MIR AGN fraction, including the MRS coordinates (astrometrically-corrected), redshift, total (8--1000$\,\mu$m) infrared luminosity, MIR AGN fraction, and available HST image data. Redshifts are measured from [Ar II] (Section 4.1). MIR AGN fractions were obtained by fitting the MRS spectra to a simple model that includes a star formation template, an AGN continuum and extinction (see Section 2 for more details). The typical error in the MIR AGN fraction is 10\% (Kirkpatrick et al.~2015)
    and the typical error on L$_{\rm{IR}}$ is 20\% (Sajina et al.~2012). The last two columns list the HST imaging band available for each source and the morphological classification from the HST data; additional details and references are provided in Section 2.1.2. 
    }
    \label{tab:my_label}
    \end{minipage}
\end{table} 

\subsection{Observations and Reduction}
\subsubsection{MIRI/MRS Data}
Observations of our targets were taken in July, August, and December of 2022 with MIRI/MRS. The observations covered the full 4.9 – 28.8 $\mu$m range using the short (A), medium (B), and long (C) subbands in all four channels. The exposure time was 2220 s in each of the three subbands. Observations of each source were reduced and assembled into spectral data cubes using the standard JWST Science Calibration Pipeline 1.11.1 (Bushouse et al.~2023) with CRDS release jwst 1100.pmap with several customizations as described in Young et al.~(2023).

\subsubsection{Ancillary HST Images}
Ancillary HST images of these sources (with the exception of FLS 3) were obtained from the Barbara A. Mikulski Archive for Space Telescopes (MAST) to examine rest-frame UV/optical morphologies. For FLS 1 and FLS 4 
we have NICMOS WFC3 FW160 (Zamojski et al.~2011), for FLS 2 and FLS 6 we have ACS WFC F475W (Lacy et al.~2007), and for COSMOS 1 we have ACS WFC F814W (Koekemoer et al.~2007). 
Previous morphological classifications are available for most of the sources in our sample, although they were derived from different studies rather than a uniform analysis. FLS 2 and FLS 6 have been classified as likely merger remnants (Lacy et al.~2007), FLS 1 as a disk-dominated, first-contact merger, and FLS 4 as a disk-dominated edge-on spiral (Zamojski et al.~2011). The morphology of COSMOS 1 has not been previously reported; based on visual inspection of the HST image, it appears relatively circular and undisturbed. The HST images are shown in Figure~\ref{fig:morph}, where they are compared with the kinematic maps.

\section{Analysis} \label{sec:analysis}
\subsection{Astrometric Corrections}
As reported in Sajkov et al.~(2024), offsets exist between these HST and MIRI/MRS datasets. To align the MIRI/MRS and HST data, we use Gaia as a common astrometric reference frame to correct for offsets between the datasets. Our MIRI/MRS data were taken concurrently with MIRI 5.6$\mu$m imaging and so we use those images to align MIRI/MRS. The astrometric shifts for each field are determined by comparing the positions of Gaia stars to their counterparts in both the MIRI and HST images. 
 

We queried the Gaia DR2 catalog (Gaia Collaboration al.~2016, Gaia Collaboration et al.~2018) to identify stars within each MIRI field. To identify the corresponding stars in the MIRI images we used the DAOPHOT algorithm implemented in Photutils (Stetson 1987). 
We calculated the difference in RA and DEC between each Gaia star and its MIRI counterpart and found the average of the RA and DEC shifts for each field were all $<0.2\arcsec$. We follow the same procedure for the HST images. FLS 4 
and COSMOS 1 did not have Gaia sources in its HST field, therefore we were only able to determine the astrometric offset in the HST fields for FLS 1, FLS 2, and FLS 6. 

With these astrometric corrections applied, we can overlay contours generated from the MRS data on the HST images (Figure \ref{fig:morph}). For one source, FLS 6, an apparent astrometric offset in declination persists even after correction. To improve the alignment between the images, we aligned the photometric center of the channel 1 ``whitelight image" (the channel 1 MRS cube summed along the spectral axis) with the photometric center of the HST image, applying an additional offset of 0.25$\arcsec$ in declination. The limited number of Gaia stars available in some fields contributes to residual uncertainties in the positions, although the alignment between HST and JWST is sufficient for the science goals in this paper. 

\subsection{Aperture Spectral Extractions}
\begin{figure}
\centering
    \begin{minipage}[t]{\textwidth}
    \centering
        \includegraphics[scale=0.75]{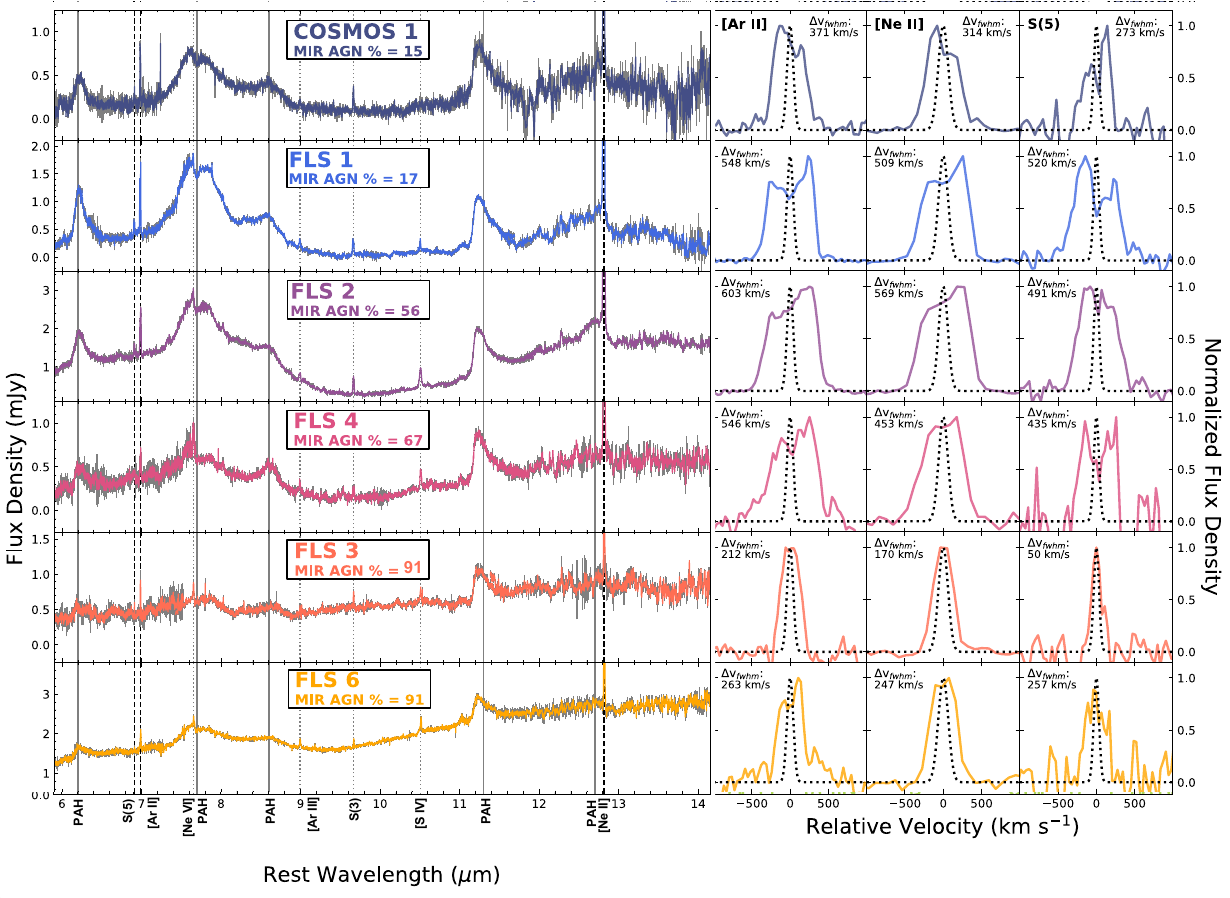}
        \caption{Left: Portion of the full spectra of the sample that includes our 3 main lines of interest extracted using a r = 0.85$\arcsec$ aperture, ordered by increasing MIR AGN fraction calculated from this data. The colored points represent the measured flux values, while the surrounding gray shading indicates the associated uncertainties. Vertical dashed black lines mark the emission lines relevant to this study: [Ar II]6.99$\mu$m, [Ne II]12.81$\mu$m and S(5)6.91$\mu$m). Vertical gray lines mark the polycyclic aromatic hydrocarbon (PAH) complexes at 6.2, 7.7, 8.6, 11.3, and 12.7 $\mu$m. Dotted gray lines mark other prominent emission lines not used in this study: [Ne VI]$\lambda$7.65$\mu$m, [Ar III]$\lambda$8.99$\mu$m, and [S IV]$\lambda$10.51$\mu$m. Right: The line profiles of a) [Ar II], b) [Ne II], and c) S(5) for o ur sample. The dotted line represents the spectral resolution element at that wavelength for reference. Low MIR AGN fraction sources (COSMOS 1, FLS 1, FLS 2, FLS4) generally show broad or double-peaked profiles, while high MIR AGN fraction (FLS 3, FLS 6)
         show narrow profiles. \label{fig:all_spec}}
    \end{minipage}
\end{figure}

To determine systemic redshifts, we extract integrated galaxy spectra from the reduced MRS data cubes using a cylindrical aperture with a radius of 0.85$\arcsec$ ($\sim$5.6 kpc), centered on the nucleus and reaching wavelengths as long as $\sim$13.5 $\mu$m (rest-frame) within the MIRI/MRS spectral range coverage. This aperture is chosen to be large enough to capture the majority of the galaxy's light across all four channels.
Given the increasing instrumental noise at longer wavelengths in Channel 4 (a known limitation of JWST MIRI/MRS), we exclude spectral lines beyond [Ne II]12.81$\mu$m rest-frame for this study.
Since this paper is only focused on velocity measurements, aperture corrections were not applied, and will be negligible for the aperture size chosen. We derive the spectral noise by extracting a noise spectrum from each data cube using a matching off-source aperture and calculating a rolling standard deviation. The extracted spectra and their uncertainty using the 0.85$\arcsec$ aperture are shown in the left panel of Figure \ref{fig:all_spec}. Here we show a portion of the full MRS spectrum that includes the main lines of interest for this paper.

For the kinematic analysis in this paper, we focus on 3 lines: two bright atomic lines tracing star formation ([Ar II]6.99$\mu$m and [Ne II]12.81$\mu$m) and the warm, pure rotational molecular hydrogen line H$_{2}$ 0-0 S(5)6.91$\mu$m, (henceforth S(5)6.91$\mu$m), chosen since it has similar resolution to [Ar II]6.99$\mu$m and is away from the 9.7$\,\mu$m silicate absorption feature. The right panel of Figure \ref{fig:all_spec} shows a zoom-in of the emission line profiles for these three lines. The other prominent lines shown in these spectra (e.g.~[Ar III], [Ne VI], [S IV], PAHs, additional H$_{2}$ lines) are not used in this study; these are featured in other papers (Young et al.~2023; Young et al.~in prep.; Yan et al.~in prep; Pope et al.~in prep).

By examining Figure \ref{fig:all_spec}, we notice that galaxies with relatively low MIR AGN fraction display lines are broad and double-peaked (COSMOS 1, FLS 1, FLS 2, FLS 4), while galaxies with higher AGN fractions have narrower emission lines (FLS 3, FLS 6). 
These line profiles, showing the integrated spectrum over the whole galaxy, motivated a detailed spatially resolved kinematic analysis described in the next few sections. 

\subsection{Kinematic Maps}
To create kinematic maps, we extracted spectra from individual spaxels (spatial pixels) using Cubeviz (JDADF Developers et al.~2024), which is tailored for JWST/MIRI data. These spectra were then used to generate velocity moment maps using a custom-built line-fitting and map-making code. Each emission line profile was modeled with a linear function for the local continuum and a single Gaussian profile centered near the expected wavelength of the line. Upon visual inspection, no broad secondary component was found in the individual spaxel analysis.

We perform this analysis on the S(5)6.91$\mu$m, [Ar II]6.99$\mu$m, [Ne II]12.81$\mu$m lines. S(5)6.91$\mu$m and [Ar II]6.99$\mu$m are bright at shorter wavelengths and thus offer good spatial resolution, while [Ne II]12.81$\mu$m is exceptionally bright and provides strong signal-to-noise, albeit with more than 1.5x lower spatial resolution. 

\begin{figure}
\centering
\begin{minipage}[t]{\textwidth}
\centering
\includegraphics[scale=0.5]{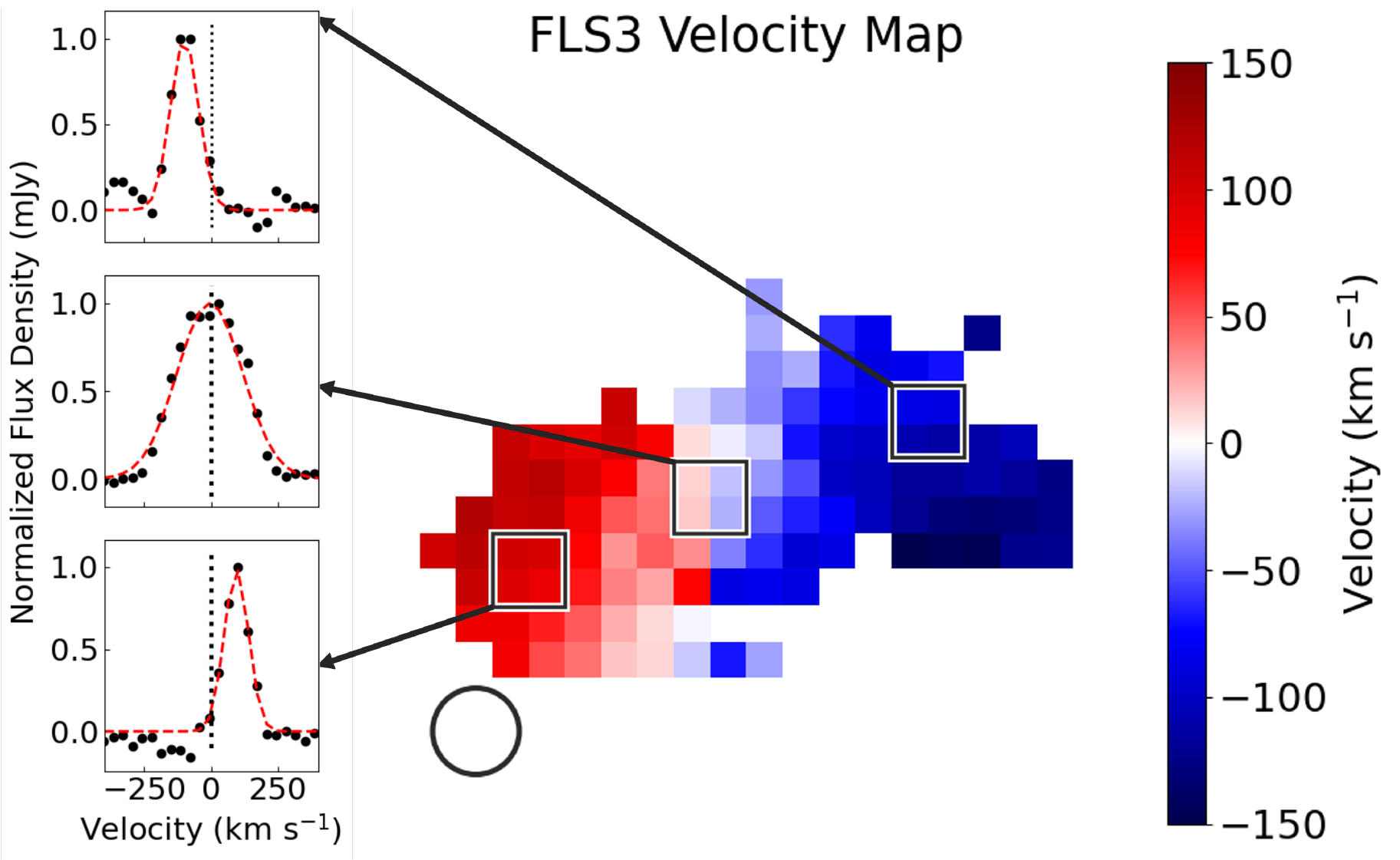}
\caption{Velocity map of FLS 3 (MIR AGN fraction = 91\% ) from the [Ar II] line. The small black circle denotes the spatial resolution at this wavelength (2.4 kpc or 0.47$\arcsec$). Shown in the inset boxes are spectra from 3 different regions (roughly the size of the spatial resolution) in the velocity map clearly showing the change in velocity across the map. An example of the Gaussian fit applied to the local continuum subtracted [Ar II] spectra is denoted by the dashed red line.}   \label{fig:inset_F3}
\end{minipage}
\end{figure}

The signal-to-noise ratio (S/N) for each spaxel is calculated by dividing the area under the Gaussian fit curve for each spaxel’s emission line by the quadrature sum of the noise over a corresponding wavelength range. These noise values are derived from the average deviation of the continuum flux on either side of the emission line for each spaxel (e.g.~Gon\c calves et al., 2010). 

We implemented a minimum threshold of S/N = 6 per spaxel, which falls in the typical values for IFU/IFS (Integral Field Unit/Integral Field Spectroscopy) studies, (F\" orster-Schreiber et al.~2009; Bellocchi et al.~2013, Wisnioski et al.~2015). Spaxels with a S/N ratio between 6 and 10 are inspected by eye to ensure the quality of the fit. For each spaxel that satisfies the S/N threshold, we calculate the velocity relative to the systemic redshift (see \S4.1 for details) to create a velocity map of each galaxy. The corresponding velocity dispersion for each spaxel is given by the standard deviation ($\sigma$) of the Gaussian fit to the emission line, which we use to construct the dispersion map. In Figure \ref{fig:inset_F3} we show an example of a velocity map generated using this process. 

\begin{figure}
\centering
\begin{minipage}[t]{\textwidth}
\centering
\includegraphics[scale=1]{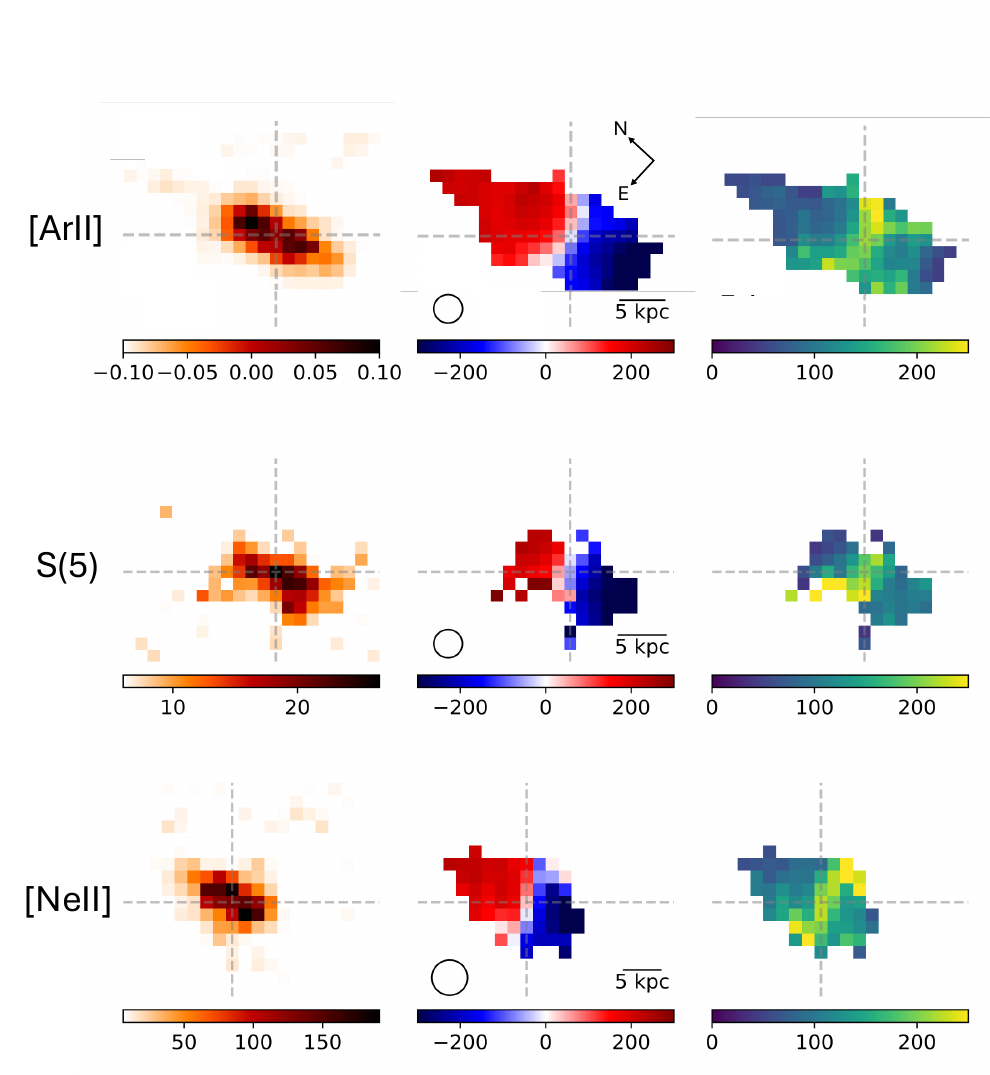}
\caption{Kinematic maps from [Ar II] (top), S(5) (middle) and [Ne II] (bottom) for FLS 1, which is a more SF-dominated galaxy (MIR AGN fraction = 17\%).
The columns show the S/N ratio, velocity, and velocity dispersion maps for each line. 
Gray dashed lines indicate the galaxy center, defined as the peak of the [Ar II]6.99$\mu$m flux.}
\label{fig:f1_kmaps}
\end{minipage}
\end{figure}

\begin{figure}
\centering
\begin{minipage}[t]{\textwidth}
\centering
\includegraphics[scale=1]{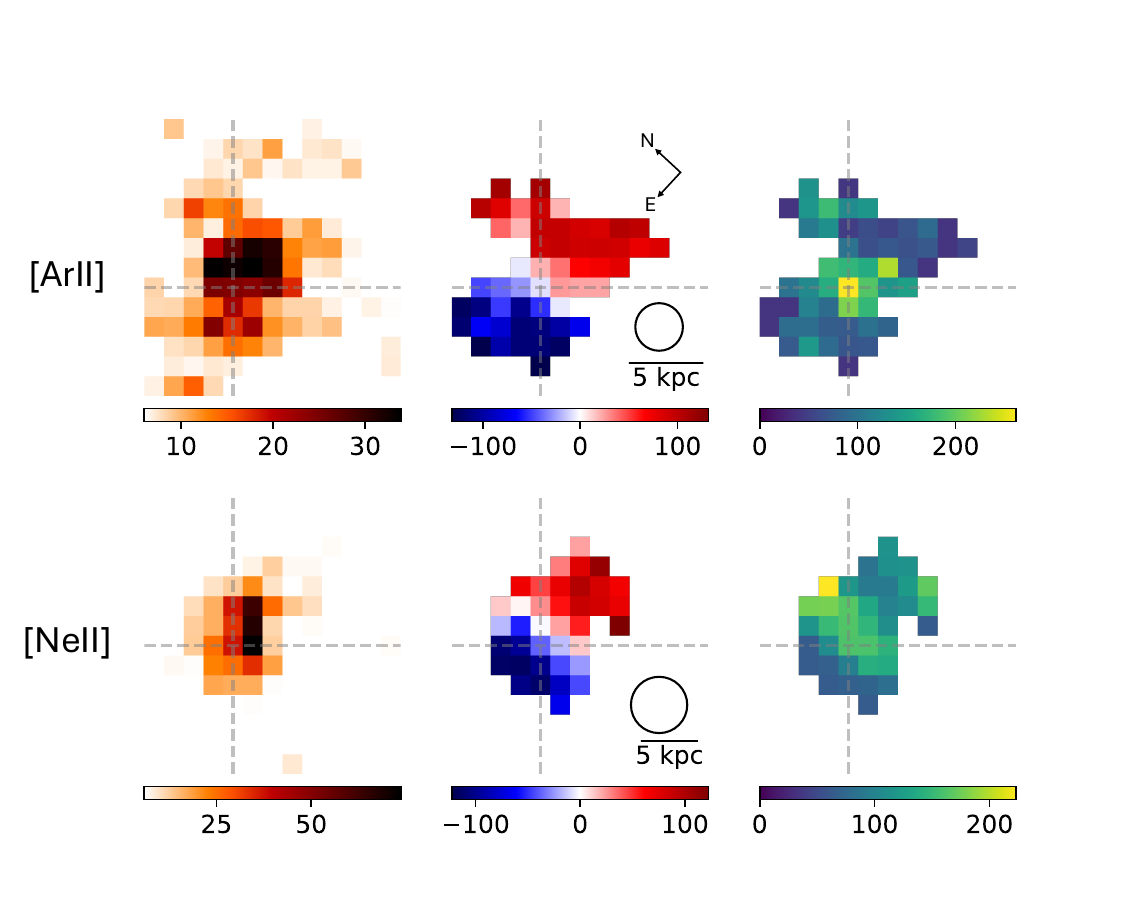}
\caption{Kinematic maps from [Ar II] (top) and [Ne II] (bottom) for FLS 6, which is a more AGN-dominated galaxy (MIR AGN fraction = 91\%).
The columns show the S/N ratio, velocity, and velocity dispersion maps for each line. 
Gray dashed lines indicate the galaxy center, defined as the peak of the [Ar II]6.99$\mu$m flux.}
\label{fig:f6_kmaps}
\end{minipage}
\end{figure}

Given the S/N threshold, we were able to generate kinematic maps in the [Ar II] and [NeII]12.81$\mu$m emission lines for all sources. For the S(5)6.91$\mu$m emission line, only COSMOS 1, FLS 1, and FLS 2 had sufficient S/N per spaxel to generate kinematic maps. In Figures \ref{fig:f1_kmaps} and \ref{fig:f6_kmaps}, we present the resulting kinematic maps for a SF-dominated source (FLS 1) and an AGN-dominated source (FLS 6), respectively. Similar figures were generated for the remaining sources (see Appendix A, Figures \ref{fig:c1_kmaps}--\ref{fig:f4_kmaps}). 

\section{Results} \label{sec:results}
\subsection{Redshifts}
The redshifts were measured for all strong emission lines using the line-fitting procedure described in the previous section using the 0.85$\arcsec$ aperture spectra (Section 3.2). We find that the redshifts for each source are consistent across the different emission lines and agree with previously determined values from Spitzer/IRS data. Although Br$\alpha$ is commonly used to measure systemic redshift, Br$\alpha$ is only detected with sufficient signal-to-noise for redshift determination in FLS 1 and FLS 2, the sources with the lowest MIR AGN fraction. 

For our sources, the two strongest atomic lines are [Ne II]12.81$\mu$m and [Ar II]6.99$\mu$m, both of which have low ionization potential indicating they are effective tracers of star-formation (e.g.~Spinoglio et al.~2015). [Ne II]12.81$\mu$m falls within channel 4, where the S/N ratio of the data degrades substantially and the spatial resolution is worse. [Ar II]6.99$\mu$m, however, was detected strongly in all sources, and resides in channel 2, which has better S/N ratio and spatial resolution. 

For this paper, we estimate the systemic redshift from the wavelength of the [Ar II]6.99$\mu$m emission line in the extracted spectrum. We convert the best-fit central wavelength to redshift, with uncertainties calculated by summing in quadrature the error from the covariance matrix of the fit and the spectral resolution at that wavelength (derived from Equation 1 in Argyriou et al.~2023). We find no systematic shifts, within the uncertainties, between the redshifts derived from different emission lines, supporting the reliability of our choice of [Ar II]6.99$\mu$m for redshift determination. The calculated redshifts from [Ar II]6.99$\mu$m are reported in column 4 of Table \ref{tab:my_label}. 

\subsection{Kinematics}
Figures \ref{fig:f1_kmaps}, \ref{fig:f6_kmaps}, and \ref{fig:c1_kmaps}--\ref{fig:f4_kmaps} show the kinematic maps for these sources. The columns display the signal-to-noise (S/N) ratio, velocity, and velocity dispersion maps, respectively, while the rows show different spectral \textbf{lines ([Ar II]6.99$\mu$m,} [Ne II]12.81$\mu$m, and S(5)6.91$\mu$m, when available). Although these maps were not corrected for beam smearing, we assess its impact in Appendix B and show that it does not affect the qualitative interpretation of the velocity and dispersion maps.
In the [Ar II]6.99$\mu$m velocity maps, most sources exhibit velocity gradients consistent with rotation, although the steepness of the gradients and the magnitude of the rotation vary across the sample. In the [Ar II]6.99$\mu$m dispersion maps, FLS 3, FLS 4 and FLS 6 show higher dispersion in the central regions compared to the rest of the galaxy. COSMOS 1 is quite compact relative to the other galaxies; its velocity map lacks a clear rotational pattern and it has higher dispersion.

While [Ne II]12.81$\mu$m is mapped for the same sources as [Ar II]6.99$\mu$m, we can only make kinematic maps of S(5)6.91$\mu$m for COSMOS 1, FLS 1, and FLS 2, which have lower MIR AGN fractions. This can be expected since S(5)6.91$\mu$m is a molecular hydrogen emission line often associated with warm, dense gas in star-forming regions. In addition, the radiation field of AGN-dominated galaxies might have photoionized the gas or photodissociated the S(5)6.91$\mu$m molecules. A comprehensive analysis of the H$_2$ lines will be presented in Yan et al.~(in prep).

Figures \ref{fig:f1_kmaps}, \ref{fig:f6_kmaps}, and \ref{fig:c1_kmaps}--\ref{fig:f4_kmaps} show that the kinematic maps are consistent across all gas tracers, with similar rotation direction and no significant line-to-line differences. Notably, the [Ne II]12.81$\mu$m dispersion maps show elevated central velocity dispersion in FLS4, FLS3, and FLS6, sources that also exhibit increased central dispersion in [Ar II]6.99$\mu$m. This suggests that the overall rotational structure is broadly similar across tracers within individual galaxies.

\subsubsection{[Ar II] Velocity and \textbf{Velocity} Dispersion Curves}
In order to quantify the kinematics in our sample, we calculate velocity and dispersion curves as a function of radius using the [Ar II]6.99$\mu$m kinematic maps, which offer sufficient S/N and the best spatial resolution in our data $\sim$0.46$\arcsec$ or $\sim$3.0 kpc). The lower SNR and spatial resolution of the [NeII] and S(5) kinematic maps limit our ability to measure velocity and dispersion radial curves reliably. We first rotate each velocity and dispersion map so that the velocity $\sim$0 axis aligns with 90$^\circ$ (i.e., horizontal), and then extract one-dimensional profiles by averaging a narrow horizontal strip of spaxels, typically $\sim$1 spatial resolution element in width (i.e., a few spaxels)—along each column. 

\begin{figure}
\centering
\begin{minipage}[t]{\textwidth}
\centering
\includegraphics[scale=0.8]{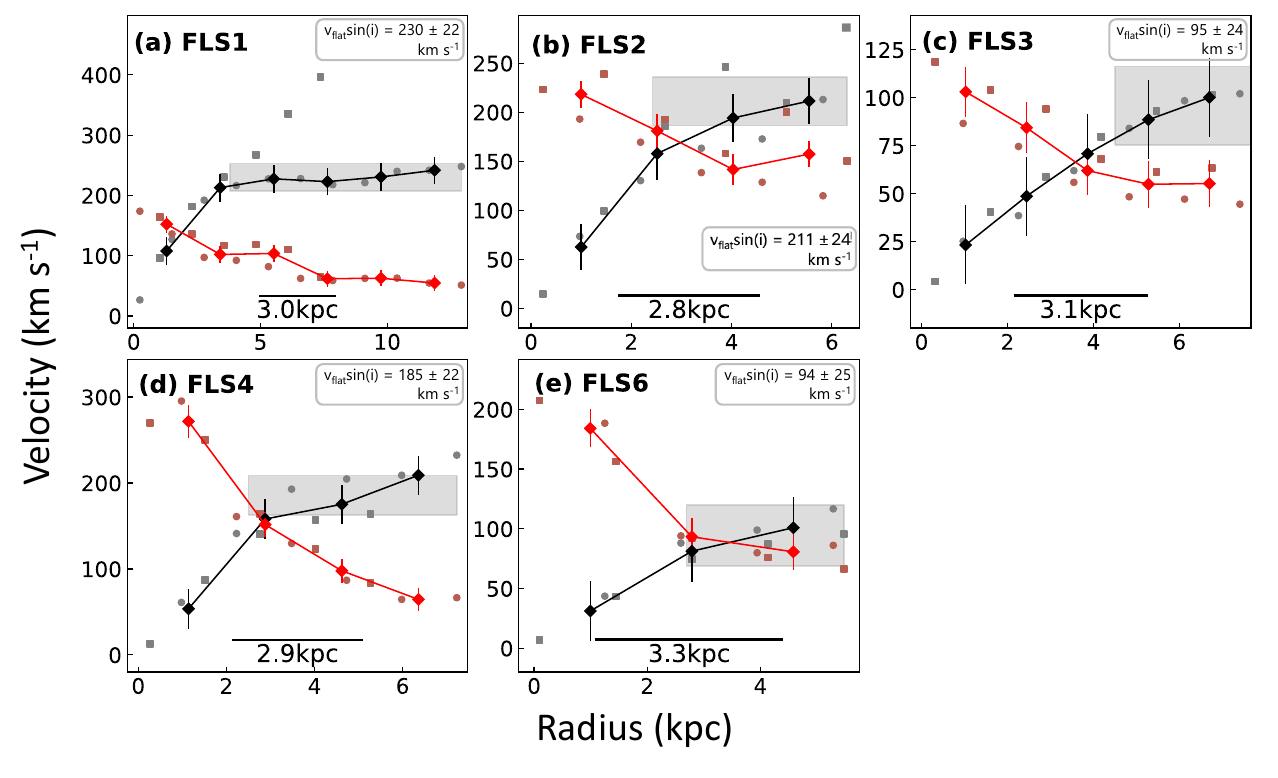}
\caption{Folded velocity (black) and velocity dispersion (red) curves. Positive velocities are marked with squares, while circles indicate negative velocities. The black scale bar represents the spatial resolution FWHM at the wavelength of [Ar II]6.99$\mu$m in physical units (kpc). The shaded region shows where we calculated v$_{\text{flat}}$sin$i$. Velocity curves shown here are not corrected for $sin(i)$. Radii (x-axes) are measured as projected distances from the galaxy center on the sky. 
\label{fig:vd_curve}}
\end{minipage}
\end{figure}

We calculate the absolute values of the velocities to produce folded velocity and dispersion curves, shown in Figure \ref{fig:vd_curve}. The galaxy center is determined from the peak of the [Ar II]6.99$\mu$m flux map (shown as the grey dashed lines in Figures \ref{fig:f1_kmaps}, \ref{fig:f6_kmaps}, \ref{fig:c1_kmaps}--\ref{fig:f4_kmaps}), which provides a well-determined central position for all sources included in this analysis. We were able to robustly measure velocity and dispersion curves for 5 out of 6 galaxies in our sample (excluding COSMOS 1 due to limited spatial extent in the velocity field). 

Although the velocity curves exhibit some asymmetries, most of our sample shows fairly regular rotation curves given the spectral resolution and measurement uncertainty. The most significantly asymmetrical velocity curve is FLS 1, where receding side of the galaxy (squares) appears to be more extended and has a higher velocity than the other side (circles). This might indicate that FLS 1 is not a perfect rotating disk and may have a disturbance on one side of the galaxy. This is consistent with findings from Young et al.~(2023), where they theorize that a recent gas accretion event may be the cause of the turbulence. 

The higher MIR AGN fraction galaxies, FLS 3, FLS 4 and FLS 6, show a steep increase ($\sim2\times$) in the dispersion towards the center, while the lower MIR AGN fraction galaxies display a shallower rise in dispersion in their central regions. We interpret this as the higher velocity dispersion being more centrally concentrated in the galaxies with higher MIR AGN fraction. While the velocity dispersion maps are not corrected for beam-smearing, we tested its effects and found that it cannot account for the central increase in dispersion observed in FLS 3, FLS 4 and FLS 6 (see Appendix B). This increased dispersion indicates higher turbulence or chaotic motion of gas in the central areas which we discuss more in Section 5.3.

\subsubsection{V/$\sigma$ as a function of AGN fraction}

\begin{table}
    \centering
    \begin{tabular}{c|cc|c|cccc}
     
        &  \multicolumn{2}{c}{Observed}&   & \multicolumn{3}{c}{Intrinsic} \\

      source &  V$_{flat}$sin(i)&  $\sigma_{obs}$ & i [best fit]  & $\sigma_{corr}$ & V$_{flat}$/$\sigma_{corr}$ & V$_{flat}$/$\sigma_{corr}$ \\
         & (km s$^{-1}$)& (km s$^{-1}$) & (degrees)& (km s$^{-1}$)&  [best fit i] & [i=90--20 deg]  \\
         \hline 
         FLS 1&  230 $\pm$ 22&   97 $\pm$ 14 & 62 &  $<$93$^{\dagger}$  & $>$2.80 & 2.47--7.23 \\
         FLS 2&  211 $\pm$ 24&   173 $\pm$ 15 & 30 &  146 $\pm$ 18 & 2.89 $\pm$ 0.48& 1.45--4.22 \\
         FLS 4&  185 $\pm$ 22&   154 $\pm$ 16 &  31 &  123 $\pm$ 20  & 2.92 $\pm$ 0.59& 1.50--4.40 \\
         FLS 3&  95 $\pm$ 24&   66 $\pm$ 12 & 26 &   $<$93$^{\dagger}$ & $>$2.33& 1.02--2.99 \\
         FLS 6&  94 $\pm$ 25&   116 $\pm$ 16 & 20 & 70 $\pm$ 27 &3.93 $\pm$ 1.84 & 1.34--3.93 \\
    \end{tabular}
    \caption{Table of kinematic measurements from [Ar II] velocity maps. Columns 2 and 3 list the observed quantities from the velocity curves in Figure \ref{fig:vd_curve}: V$_{flat}$sin(i) and $\sigma_{obs}$, the average velocity dispersion measured over all radii. Column 4 shows the best fit $i$ value from the method outlined in section 4.2.2. Columns 5-7 list the intrinsic quantities. Column 5 is the velocity dispersion corrected for the instrument resolution. Column 6 shows the intrinsic V$_{flat}$/$\sigma_{corr}$ assuming the best fit inclination in column 4. Column 7 lists the range of V$_{flat}$/$\sigma_{corr}$ could span assuming the inclination ranges from 90--20 degrees. \\
    $^{\dagger}$ $\sigma_{obs}$ is unresolved for FLS 3 and marginally resolved for FLS 1. For these two sources we set $\sigma_{corr}$ equal to the instrumental resolution (93 km s$^{-1}$) which will be an upper limit to the true intrinsic velocity dispersion. As a result, the V$_{flat}$/$\sigma_{corr}$ values for these two sources can be considered a lower limit.
    }
    \label{vflat_tab}
\end{table}

We can use the velocity and dispersion curves, and the assumed inclination, to estimate the intrinsic $V/\sigma$ to determine if the rotation is stable.
$V_{flat}\sin(i)$ (the observed velocity) is estimated from the velocity curves as the average over the shaded region in Figure \ref{fig:vsig_AGN}, while $\sigma_{obs}$ is estimated as the average of the velocity dispersion curve at all radii, conservatively including the higher dispersion central region\footnote{Calculating $\sigma_{obs}$ over just the outer radii, where $V_{flat}\sin(i)$ is defined, would result in a lower value (spectrally unresolved for most sources) and thus a higher value of $V/\sigma$.}. We correct $\sigma_{obs}$ for the instrumental resolution, which is $\sim$93 km s$^{-1}$ at the observed wavelengths of [Ar II] 6.99 $\mu$m for our sample. In order to calculate the intrinsic $V_{flat}/\sigma_{corr}$, we estimate the inclination by fitting the white-light images of the galaxies with a 2D Gaussian. The ratio of the full width at half maximum (FWHM) in each dimension is used as the axis ratio of the galaxy (\( b/a \)), which is then used to calculate the inclination (\( i = \cos^{-1}(b/a) \)). Given the uncertainty in estimating the inclination, we also explore the inferred $V/\sigma$ over the full range of $i$ from 20--90 degrees. We use the inclination to calculate $V_{flat}$ and combine this with the average intrinsic dispersion $\sigma_{corr}$ to estimate $V_{flat}/\sigma_{corr}$. All relevant quantities derived from the velocity maps are listed in Table \ref{vflat_tab}.

We note that for one of our sources, FLS 3, $\sigma_{obs}$ is unresolved, while for another of our sources, FLS 1, $\sigma_{obs}$ is only marginally resolved. Therefore, for these two sources, we set $\sigma_{corr}$ to the spectral resolution, as an upper limit, and the calculated $V/\sigma$ values will be lower limits. Consequently, any kinematic interpretations for these sources should be considered tentative.

The V/$\sigma$ ratio is a commonly used diagnostic for assessing the dynamical state of galaxies, where higher values typically indicate rotation-supported systems and lower values suggest pressure-supported or dispersion-dominated dynamics (e.g., Weiner et al.~2006; Gon\c calves et al.~2010). While a threshold of V/$\sigma$ $\simeq$ 1 is often adopted as a rough dividing line between these regimes (Weiner et al.~2006), this value is somewhat arbitrary and varies across studies (e.g., Gon\c calves et al.~2010), particularly as typical values in low-redshift disk galaxies are significantly higher (V/$\sigma$ $\simeq$ 10–20; F\" orster-Schreiber et al.~2009; Wisnioski et al.~2015). Local star-forming galaxies have velocity dispersions of $\sim$25 km s$^{-1}$, while local LIRGs have velocity dispersions that range from 30--100 km s$^{-1}$ and it is thought that star formation plays a relatively small role in driving the higher velocity dispersion in local LIRGs (Arribas et al.~2014). 

Figure~\ref{fig:vsig_AGN} shows V/$\sigma$ as a function of MIR AGN Fraction. All galaxies in our sample exhibit intrinsic $V/\sigma$ values greater than 1, consistent with rotationally supported galaxies (Weiner et al.~2006 and references within). This is expected given the sample’s selection criteria: 
massive, IR-luminous systems with substantial dust and gas content. These properties are more characteristic of extended, star-forming disk galaxies than of dispersion-dominated systems. Consequently, the observed prevalence of rotational support likely reflects both the physical nature of these systems and the selection bias toward dynamically mature, disk-dominated galaxies at this epoch.


It is notable that even sources with high MIR AGN fractions maintain ordered kinematics, suggesting that moderate to high AGN activity does not strongly disrupt the overall gaseous disk structure in these systems. By examining the velocity dispersion maps (Figures \ref{fig:f1_kmaps}, \ref{fig:f6_kmaps}, \ref{fig:c1_kmaps}--\ref{fig:f4_kmaps}) and the radial dispersion profiles (Figure \ref{fig:vd_curve}), we find that the velocity dispersion is more centrally concentrated in the higher MIR AGN fraction sources (FLS 3 and 6). At the same time, the galaxy-integrated spectra of these systems show relatively narrow emission-line profiles (Figure 1, right panel). This behavior likely reflects differences in the spatial extent of the line-emitting regions. In AGN-dominated galaxies, the line emission is more compact and centrally concentrated, with the nuclear component dominating the integrated spectrum and outshining the host galaxy disk emission. In contrast, the SF-dominated (low AGN fraction) systems exhibit double-peaked emission-line profiles, consistent with rotation-dominated kinematics extending over several kiloparsecs.

While the AGN fraction reflects the relative contribution of AGN emission compared to star formation, AGN luminosity provides a measure of the absolute energy output available to influence the host galaxy’s kinematics. One might therefore expect a clearer correlation between AGN luminosity and $V/\sigma$ if AGN-driven feedback plays a significant dynamical role. However, repeating this analysis with $V/\sigma$ as a function of AGN luminosity similarly revealed no clear trend. 

\subsubsection{Constraints on the Dynamic Masses}
\begin{figure}
\centering
\begin{minipage}[t]{\textwidth}
\centering
\includegraphics[scale=0.7]{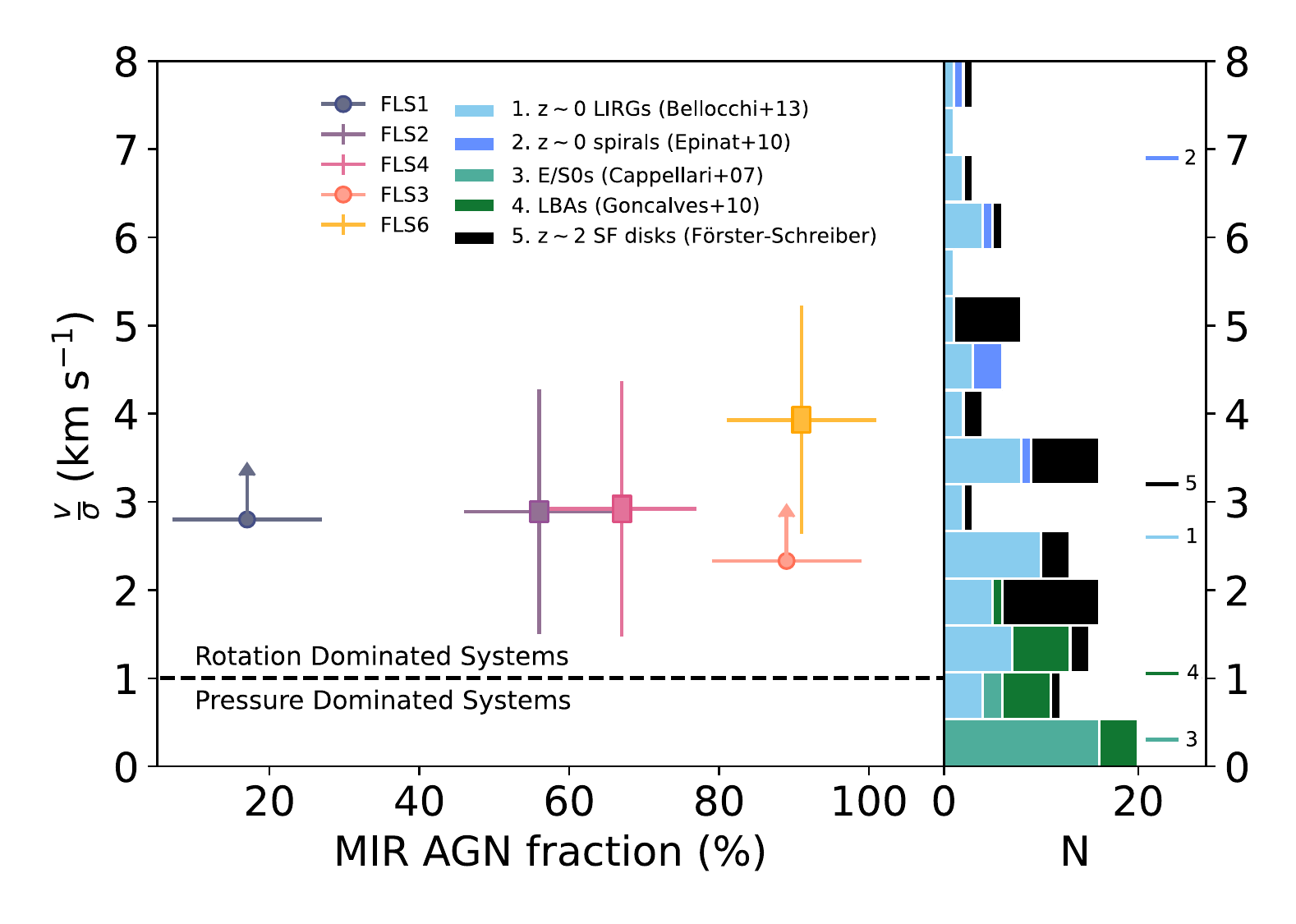}
\caption{Intrinsic \( V/\sigma \) for our $z\sim0.5$--0.6 LIRGs as a function of MIR AGN fraction. The rectangular markers designate the inclination-corrected \( V/\sigma \) using the best-fit i, while their vertical axis length reflects the uncertainty given this value of i. The circular markers denote lower limits, where $\sigma$ is marginally resolved or unresolved. Error bars indicate the full range of \( V/\sigma \) from a range of possible inclination values (Table 2, last column). Gray lines represent the median of \( V/\sigma \) for comparison galaxy samples (see legend for different populations). The dashed black line represents \( V/\sigma \) = 1, which is used to distinguish between rotation and pressure dominated systems. FLS 3 is slightly offset in the x-direction for clarity, as it overlaps with FLS 6 (both have MIR AGN fraction = 91\%)}.
\label{fig:vsig_AGN}
\end{minipage}
\end{figure}

We can use the measured velocities to estimate the dynamical masses of these galaxies. Given the large uncertainty in the dynamical masses due to the assumed inclination, our goal for the dynamical masses is to verify that the values are reasonable for these types of galaxies. We calculate the dynamical masses ($M_{dyn}$) using the standard formula,
\begin{equation}
    M(R)_{dyn} = \frac{V(R)^2R}{G}.
\end{equation} 
We define $R$ as the turnover radius of the rotation curve, typically between 1 and 3.5 kpc as inferred from their velocity curve (Figure 5), and use the velocity at this radius for $V(R)sin(i)$. Since estimating the dynamical masses of these galaxies is strongly influenced by their inclination (which is highly uncertain), we calculate the range of dynamical masses considering the full range of inclination from 20--90 degrees. This results in dynamical masses for these galaxies ranging from 10$^{10}$--10$^{11}$ $M_\odot$, similar to the range dynamical masses of our comparison samples (10$^{9}$--10$^{11}$ $M_\odot$) shown Figure \ref{fig:vsig_AGN}. These dynamical masses are consistent with massive, rotationally supported systems typical of LIRGs at $z \sim 0.5$–$0.6$, where galaxy evolution processes such as quenching or AGN feedback may become more significant (Madau \& Dickinson 2014 and references therein).

While they share kinematic characteristics with local Seyfert galaxies, such as ordered disk rotation and central velocity dispersion enhancements, their higher infrared luminosities and redshifts suggest they represent a more dynamically active phase of galaxy evolution. This is consistent with their AGN luminosities (10$^{10}$–10$^{12}$ L$_\odot$), which are indicative of moderate to high AGN power—higher than typically seen in local Seyfert galaxies. 

\section{Discussion} \label{sec:discussion}
\subsection{Rotation Dominated LIRGs at $z = 0.5$--$0.6$}
From Figure \ref{fig:vsig_AGN}, we observe that all galaxies in our sample exhibit \( V/\sigma \) values of \( \geq 1 \), indicating that they are predominantly rotation-dominated systems. The presence of ordered rotation in our sample indicates that these are massive, mature systems, suggesting that these galaxies have not recently experienced a major merger.  This is broadly consistent with expectations from cosmological simulations such as those by Hopkins et al.~(2009), which predict that many massive galaxies can retain or reform rotational support following earlier merger activity. When compared to the literature samples shown in the right panel of Figure \ref{fig:vsig_AGN}, our galaxies exhibit $V/\sigma$ values of $\sim$ 2.5--4, similar to those measured for local LIRGs (Bellocchi et al.~2013) and $z\sim2$ star-forming disks \textbf{(F\"orster-Schreiber et al.~2018)}. In contrast, these values are significantly lower than those of local spiral galaxies (Epinat et al.~2010), which typically show much higher $V/\sigma$ values of $\sim$ 7. This comparison suggests that while our systems are clearly rotation-dominated, they remain dynamically hotter than present-day spirals.

Mergers typically introduce significant turbulence and random motions within galaxies. These disruptive events can lead to chaotic kinematic states, often transforming galaxies into pressure-supported systems, particularly in AGN-dominated elliptical galaxies (Oh et al.~2022, and references therein). In AGN-dominated elliptical galaxies, quasar-mode feedback can drive large-scale outflows impacting the entire host galaxy. However, feedback from Seyfert-like AGNs often remains confined to the nuclear regions. While AGN luminosity appears to be a key factor in determining the spatial scale of host galaxy impact, obscuration and host morphology likely modulate this effect. Quantitatively, AGN with $L_{bol} \geq 10^{45}$ erg/s are typically associated with host-wide outflows, while those below this threshold often influence only the central few hundred parsecs. Further disentangling these effects would require spatially resolved multiwavelength AGN diagnostics, particularly in obscured, interacting systems like LIRGs, which can be explored in future works. While the kinematic properties of these IR-luminous galaxies seem broadly similar to those of local Seyfert galaxies (e.g., disk-like rotation with central dispersion enhancements), our sources are more IR-luminous and lie at z $\sim$ 0.5–0.6, a time when galaxies are typically more gas-rich.

Furthermore, much of our analysis focuses on the kinematics of ionized gas, specifically the [Ar II]6.99$\mu$m and [Ne II]12.81$\mu$m emission lines. These kinematics can vary depending on the dominant ionization source. Gas ionized by star formation typically exhibits a motion (ordered, rotational) different than gas ionized by an AGN (non-circular and/or disturbed near nucleus) or by shocks (irregular, with local deviations from galactic rotation). The connection between ionized gas kinematics and ionization source has been highlighted in prior work (e.g., Oh et al.~2022 and references therein), and is particularly relevant for our sample of IR-luminous galaxies, which likely host a combination of star formation, AGN activity, and possibly shocks from outflows or mergers. In such cases, the motion of the ionized gas can itself serve as a diagnostic of the prevailing ionization mechanisms. For our sample, we might expect to observe a trend where rotational motion decreases as the MIR AGN fraction increases. All of our sources are rotation-dominated and the best fit \( V/\sigma \) does not show any trend with AGN fraction. However, considering the uncertainty in the inclination (and thus the uncertainties associated with the \( V/\sigma \) ratio, see Figure \ref{fig:vsig_AGN}), we are unable to rule out any trend with AGN fraction. For the sources where $\sigma$ is  unresolved (FLS 3) or marginally resolved (FLS 1), the inferred $V/\sigma$ values represent lower limits. Combined with the uncertainty in the inclination ($i$) and the modest sample size, this introduces additional uncertainty in the kinematic interpretation of these galaxies. Nevertheless, this analysis highlights the capability of MIRI/MRS to probe the kinematics of these systems, and a larger sample will enable more robust constraints as additional observations of U/LIRGs at these redshifts become available.

\subsection{Comparing the Kinematics to the Optical Morphology}
\begin{figure}
\centering
\begin{minipage}[t]{\textwidth}
\centering
\includegraphics[width=\textwidth]{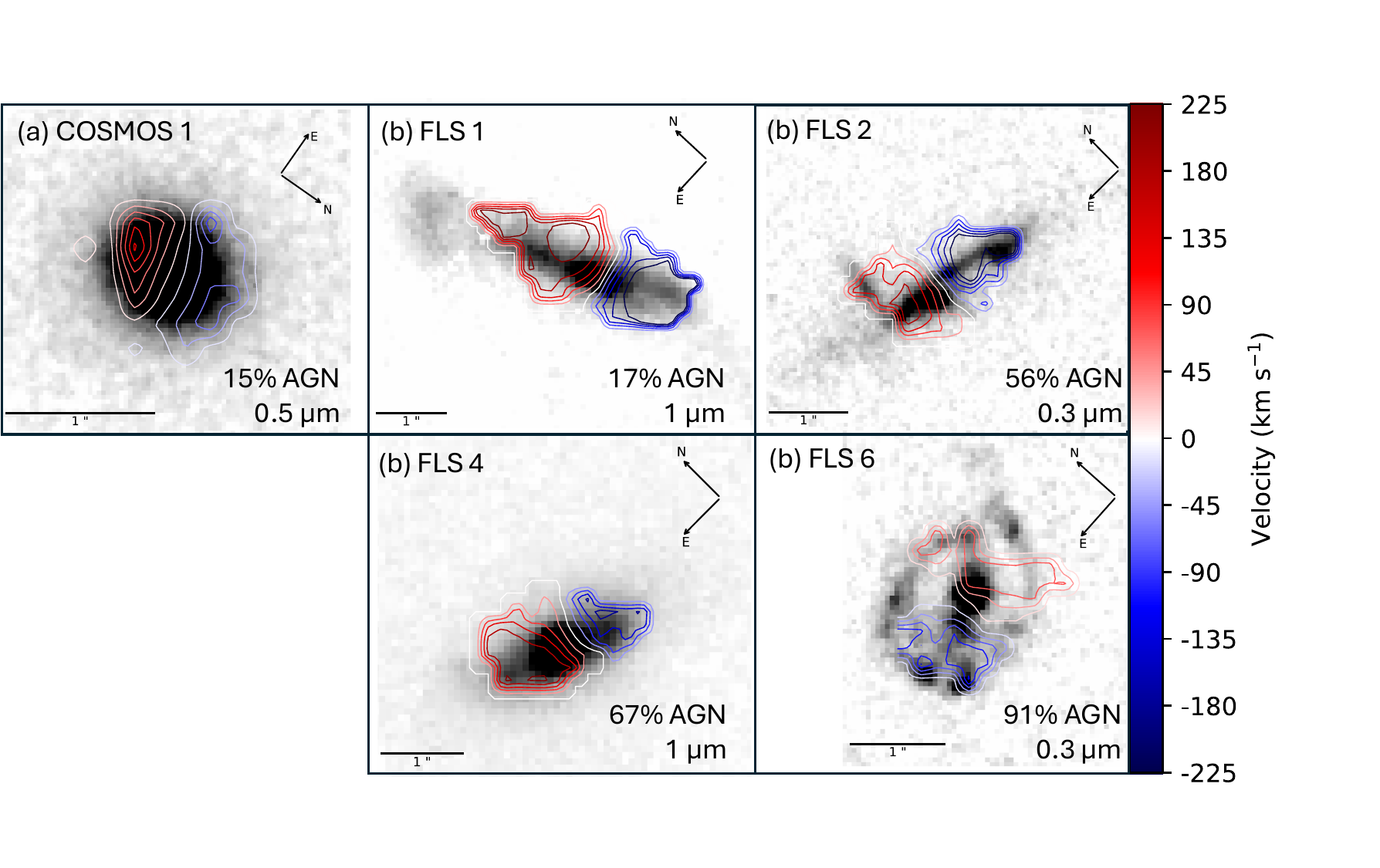}   
\caption{HST images overlaid with [Ar II] velocity map contours for five of our sources (labeled in the top left corner). The rest wavelength for the HST image is given in the bottom right corner for each source. To facilitate direct comparison with the MIRI kinematic maps, the HST images in this figure have been rotated to the native MIRI frame. 
\label{fig:morph}}
\end{minipage}
\end{figure}

UV/optical morphology, which traces the distribution of star light, is another way to assess a galaxy's evolutionary state, with clues as to whether it is undergoing a merger or resembles a stable disk. For our sources, we use archival HST NICMOS F160W, HST ACS F814W, or ACS WFC F475W imaging (probing the rest-frame UV/optical light) to examine the distribution of stellar light in relation to the gas motion observed with MIRI. Figure \ref{fig:morph} shows velocity contours derived from our [Ar II]6.99$\mu$m velocity maps overlaid onto Hubble Space Telescope (HST) UV/optical images of the sources. The available HST imaging is heterogeneous: some bands probe light blueward of the Balmer break and others redward, sampling different stellar populations (younger vs. older). This heterogeneity precludes a systematic, uniform assessment of the optical morphologies across the full sample. 

From Figure \ref{fig:morph}, we can see that the images of FLS 1, FLS 2, and FLS 6 show clumpy and/or disturbed morphology, which could suggest recent star formation or interactions, although dust obscuration is likely biasing the visual interpretation of the morphology of FLS 2 and FLS 6 as they are rest-frame NUV images. Meanwhile, other galaxies (e.g.~COSMOS 1 and FLS 4) have smooth morphologies in the optical/NIR, which is sensitive to older stellar populations. The variation in morphology across different wavelengths, combined with the effects of dust and the presence of diverse stellar populations makes it difficult to distinguish spheroids from disks based on morphology alone. Despite these morphological differences, the ionized gas kinematics of these galaxies tell us that they are all rotating disks. Among local LIRGs, morphological analysis suggest the bulk are undergoing mergers/interactions and only $\sim$35\% are isolated undisturbed disks (Stierwalt et al.~2014). However kinematic analysis suggests the vast majority are rotating disks with high V/sigma (Bellochi et al.~2013). This underlines how important it is to combine kinematic and morphological diagnostics to fully characterize their evolutionary state.

\subsection{Increased Velocity Dispersion in FLS 3,4,6}
While large-scale velocity fields trace the dominant motions of a galaxy's ionized gas, velocity dispersion fields probe the level of turbulent or random motions within the gas. Using the kinematic maps, we can also compare different regions within each galaxy to identify localized enhancements in dispersion, potentially giving us hints about the source of ionization. As illustrated in Figures~\ref{fig:f1_kmaps} \textbf{and~\ref{fig:f6_kmaps} 
(see} also Figures~\ref{fig:c1_kmaps} -~\ref{fig:f4_kmaps}), the higher MIR AGN fraction galaxies (FLS 3 and FLS 6) clearly exhibit increased velocity dispersion in their central regions. Feedback from an AGN, such as radiation pressure and AGN-driven winds, can disrupt the interstellar medium (ISM) throughout the host galaxy (e.g., Nesvadba et al.~2011; Veilleux et al.~2020; Fluetsch et al.~2021). If strong outflows were present and superimposed on a rotating disk, we would expect to observe a broad or blueshifted component in the line profiles; features that are not seen in our data (e.g., Figure \ref{fig:all_spec}). This absence may be due to the spatial resolution limitations of our observations, which may not fully resolve the nuclear regions where such outflows are typically most prominent. The lack of significant disruption to disk dynamics, despite high MIR AGN fractions, may suggest that the AGN in these systems are turning on in already mature, rotationally supported galaxies, where the disk structure is stable and more resistant to disruption.

Since recent major mergers are unlikely given the stable rotation we observe, it is possible that these AGN systems are evolving primarily through secular processes or minor interactions. This raises the question of what mechanisms are driving both star formation and AGN activity in the absence of mergers. These galaxies may be experiencing cold accretion or smooth gas inflow, which can supply both the supermassive black hole and the star-forming disk without triggering significant structural disruption (Martin et al.~2018; Wang et al.~2020). In this scenario, the AGN could be fueled by a steady inflow of cold gas, maintaining a rotation-dominated structure while simultaneously supporting central supermassive black hole growth. We do not suggest that the current disk structure necessarily formed via IGM accretion; rather, such accretion could be sustaining or rejuvenating gas inflow into the central regions of otherwise settled disk galaxies. Simulations show that filamentary cold gas can reach kpc or even sub-kpc scales, particularly at intermediate redshifts, and may intermittently trigger AGN activity without disturbing the global rotation-dominated dynamics (Dekel et al.~2009). The concurrent suppression of S(5)6.91$\mu$m emission in these systems, potentially due to photodissociation or ionization of H$_2$ by the AGN radiation field, further supports the idea that AGN feedback is active but remains confined to the central regions, altering the gas reservoir without disturbing the global disk (Ogle et al.~2024). Simulations show that minor mergers or weak tidal interactions may also contribute to this process (Moster et al.~2010). 

Other secular processes such as bar formation and the gradual buildup of a stellar bulge may also be contributing to the high central velocity dispersion of these galaxies (Kormendy et al.~2009; Michtchenko et al.~2018). Future NIRCam imaging could help test for the presence of bars in these systems, as has been successfully done for highly obscured submillimeter galaxies (SMGs), where deep NIR observations revealed otherwise hidden bar structures (McKinney et al.~2024 and references within). These galaxies could also be experiencing concurrent central starbursts including stellar winds and supernova explosions which can also increase the velocity dispersion (Ostriker \& Shetty~2011, Martizzi~2019).

Since these systems are currently dominated by rotation-supported ionized gas disks, a major merger likely occurred at least 400--800 Myr ago (e.g., Lotz et al.~2008; Hopkins et al.~2009; Hung et al.~2016), allowing time for dynamical relaxation. In contrast, the typical duty cycle of a moderate-luminosity AGN ($L_{bol}$ $\sim$ $10^{43}$--$10^{44.5}$ erg $s^{-1}$) is estimated at $\sim$ 10 – 100 Myr (e.g., Schawinski et al.~2015; Hickox et al.~2014), suggesting that AGN activity may not be directly linked to the most recent large-scale dynamical disturbance. 

\section{Summary \& Conclusion} \label{sec:conclusion}

In this study, we analyzed a sample of five LIRGs and one ULIRG observed at redshift $z$ = 0.5--0.6 using MIRI/MRS on JWST, as part of the Halfway to the Peak Program. This redshift corresponds to a period of heightened cosmic star formation activity compared to the local universe, providing an ideal opportunity to investigate the complex interplay between star formation, AGN activity, and gas dynamics during a critical phase of galaxy evolution. By spanning a range of MIR AGN fractions (from 15\% to 97\%), our sample allows us to explore how AGN activity may influence the kinematic signatures observed in these galaxies.  
We explored the line profiles, velocity and dispersion maps for the [Ar II]6.99$\mu$m, [Ne II]12.81$\mu$m and S(5)6.91$\mu$m spectral lines, tracing the distribution and kinematics of different gas components. 
This is the first time MIR atomic lines are used to probe the kinematics of galaxies outside the local Universe. The main findings of this study are:
\begin{enumerate}
    \item Ordered rotation in all sources: All galaxies in the sample display clear signs of ordered rotation in the velocity maps and have V/$\sigma$ ratios values consistently above 1, regardless of their MIR AGN fraction. This suggests that these galaxies represent mature disks that have already settled into rotational support.
    
    \item Consistent kinematics across gas species: The consistent kinematics observed across [Ar II]6.99$\mu$m, [Ne II]12.81$\mu$m and S(5)6.91$\mu$m suggest that these lines trace the same underlying dynamical structure. This strengthens the case for using [Ar II]6.99$\mu$m as a reliable star formation tracer in the JWST era, particularly at higher redshifts where other lines may be inaccessible.

    \item Morphology and kinematics: Comparing the UV/optical morphology and gas kinematics reveal that while NUV images for some sources suggest clumpy, interaction-driven structures and optical/NIR images for other sources appear smooth and ambiguous, ionized gas kinematics confirm that all these galaxies are rotating disks, demonstrating the power of using both morphological and kinematic characterization to determine the evolutionary state of the galaxy.  
    
    \item Increased central velocity dispersion: From the velocity dispersion maps and profiles, we find elevated central velocity dispersions in at least half of our sample, with significantly higher values ($\sim2\times$) in sources with higher MIR AGN fractions. This, combined with the magnitude of the dispersion and the suppression of S(5)6.91$\mu$m in these galaxies, suggests that this phenomenon is likely driven by energy injection from the AGN. However, we note that alternative explanations, such as a stellar bulge/bar or feedback from stellar winds and supernovae, could also contribute to the observed increase in central dispersion. 
 
\end{enumerate}

These findings suggest that the overall kinematic structure of the gaseous galactic disk remains ordered in these z $\sim$ 0.55 IR-luminous galaxies, even in the presence of significant AGN activity. Galaxies with higher MIR AGN fractions do exhibit elevated central velocity dispersions, but the disks as a whole appear dynamically mature and rotationally supported. This supports the interpretation that these systems have already assembled substantial stellar and gas mass, and that AGN likely turned on after disk formation. In this evolutionary stage, feedback may operate in a more intermittent or localized fashion, insufficient to disrupt global disk stability or significantly suppress star formation. These results are consistent with previous studies showing that AGN feedback does not necessarily disturb disk kinematics in massive, evolved systems. Instead, they may be evolving primarily through secular processes, quenching from the inside out as the supply of cold gas is gradually exhausted. Our results highlight a potential transitional regime in galaxy evolution where AGN activity and disk stability coexist, offering insight into the timing and conditions under which feedback has the greatest impact. Future observational studies, with a larger sample observed in similar mid-infrared tracers, will be necessary to establish the nature of this population.

\section*{Acknowledgments}

We thank the referee for their careful review and constructive feedback, which improved the clarity of this work. We acknowledge that the University of Massachusetts Amherst (UMass Amherst), where this work was conducted, is located on the unceded ancestral lands of the Pocumtuc Nation, in the Norrwutuck region along the Kwinitekw (Connecticut River). We recognize and honor the ongoing presence and stewardship of Indigenous peoples. As a Land Grant institution, UMass Amherst also occupies land obtained through the Morrill Act of 1862, which expropriated Indigenous territories to support the development of public universities. We recognize this historical context and the responsibilities it entails.

We also wish to acknowledge the many individuals whose labor sustains the institutions that make this research possible, including administrative, technical, facility maintenance, and janitorial staff at UMass Amherst, STScI, and affiliated organizations, whose contributions are often unrecognized but essential to the functioning of the scientific enterprise. Finally, we acknowledge the programs that support equitable access and participation in higher education and research, without which this work would not have been possible.

This paper makes use of data from observations by NASA/ESA/CSA JWST obtained at the Space Telescope Science Institute, which is operated by the Association of Universities for Research in Astronomy, Incorporated, under NASA contract NAS5-03127. Support for program number JWST-GO-01762 was provided through a grant from STScI under the NASA contract NAS5-03127.

\appendix
\section{Kinematic Maps}
This appendix includes the [ArII], [Ne II]12.81$\mu$m, and S(5)6.91$\mu$m kinematic maps for COSMOS 1, FLS 2, FLS 3, FLS 4; similar to Figures \ref{fig:f1_kmaps} and \ref{fig:f6_kmaps} in Section 4.2.

\begin{figure}
\centering
\begin{minipage}[t]{\textwidth}
\centering
\includegraphics[width=\linewidth]{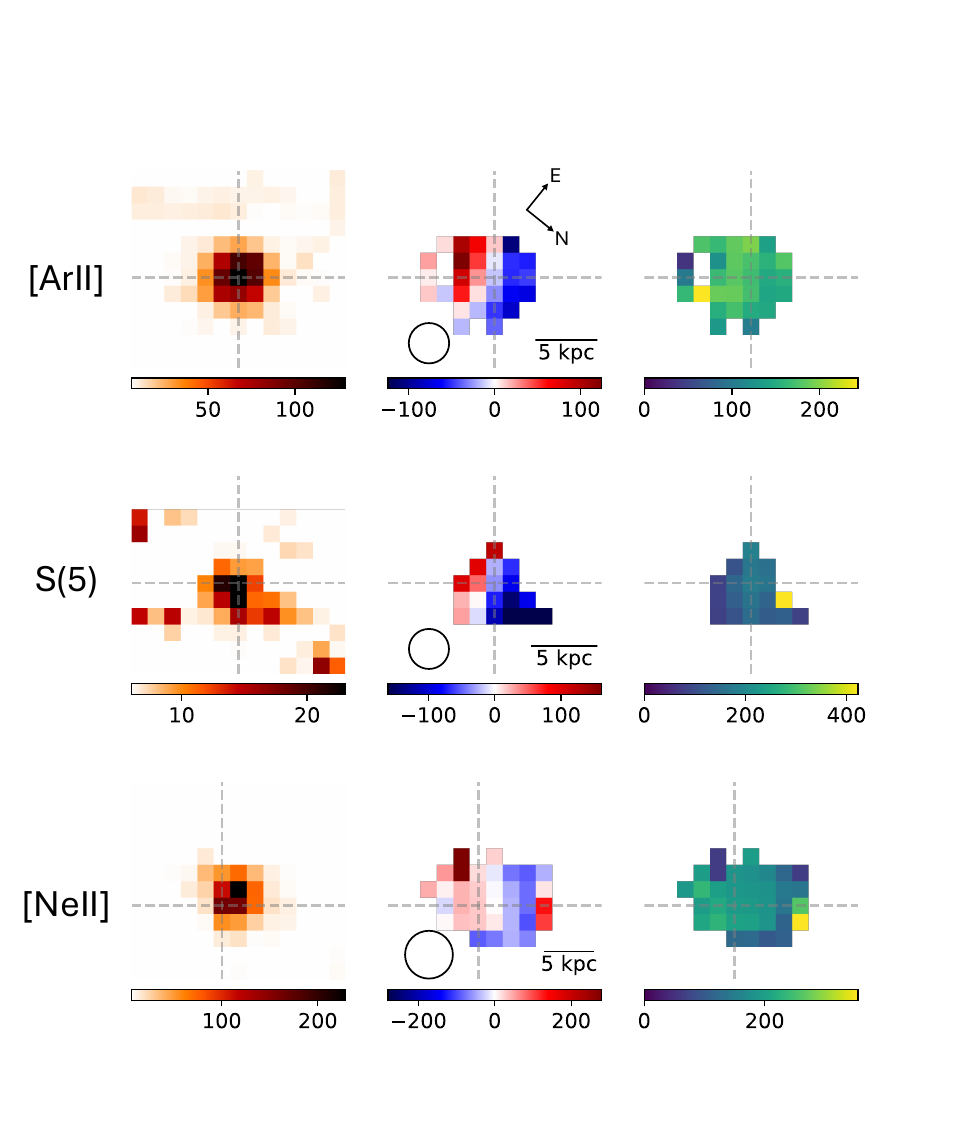}
\caption{Comparison of the kinematic maps from [Ar II] (top), S(5) (middle) and [Ne II] (bottom) for COSMOS 1. Gray dashed lines indicate the galaxy center, defined as the peak of the [Ar II]6.99$\mu$m flux.
\label{fig:c1_kmaps}}
\end{minipage}
\end{figure}

\begin{figure}
\centering
\begin{minipage}[t]{\textwidth}
\centering
\includegraphics[width=\linewidth]{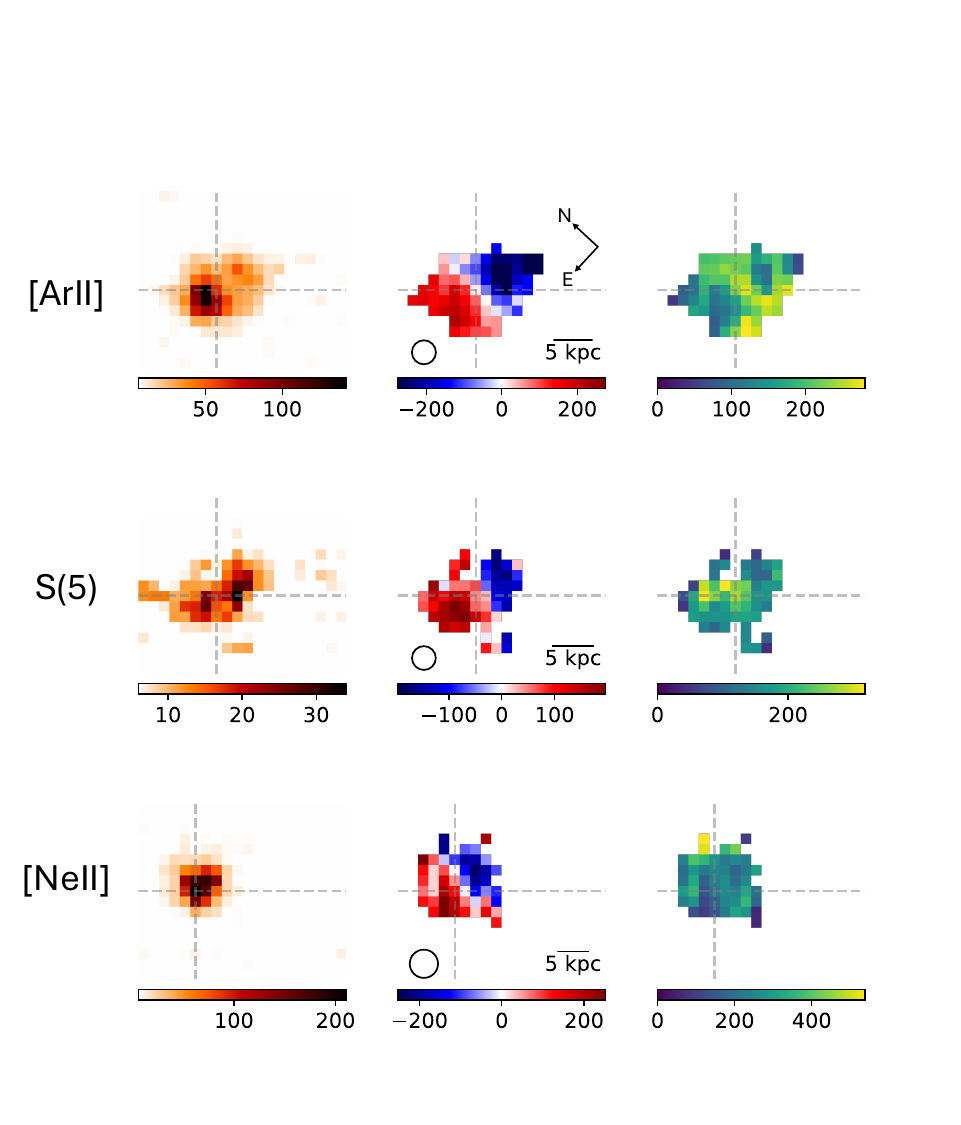}
\caption{Comparison of the kinematic maps from [Ar II] (top), S(5) (middle) and [Ne II] (bottom) for FLS 2.
The columns show the S/N ratio, velocity, and velocity dispersion maps for each line. Gray dashed lines indicate the galaxy center, defined as the peak of the [Ar II]6.99$\mu$m flux.
\label{fig:f2_kmaps}}
\end{minipage}
\end{figure}

\begin{figure}
\centering
\begin{minipage}[t]{\textwidth}
\centering
\includegraphics[width=\linewidth]{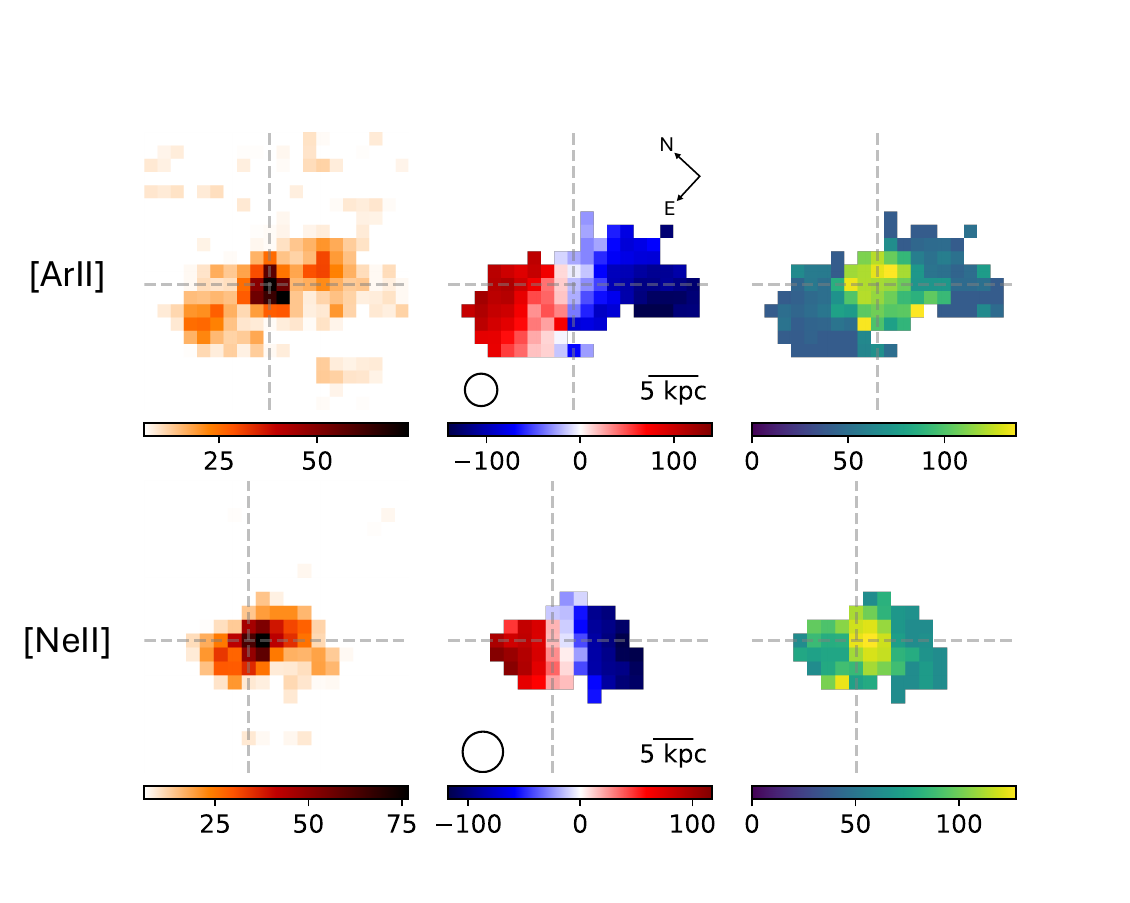}
\caption{Comparison of the kinematic maps from [Ar II] (top), S(5) (middle) and [Ne II] (bottom) for FLS 3. The columns show the S/N ratio, velocity, and velocity dispersion maps for each line. Gray dashed lines indicate the galaxy center, defined as the peak of the [Ar II]6.99$\mu$m flux.
\label{fig:f3_kmaps}}
\end{minipage}
\end{figure}

\begin{figure}
\centering
\begin{minipage}[t]{\textwidth}
\centering
\includegraphics[width=\linewidth]{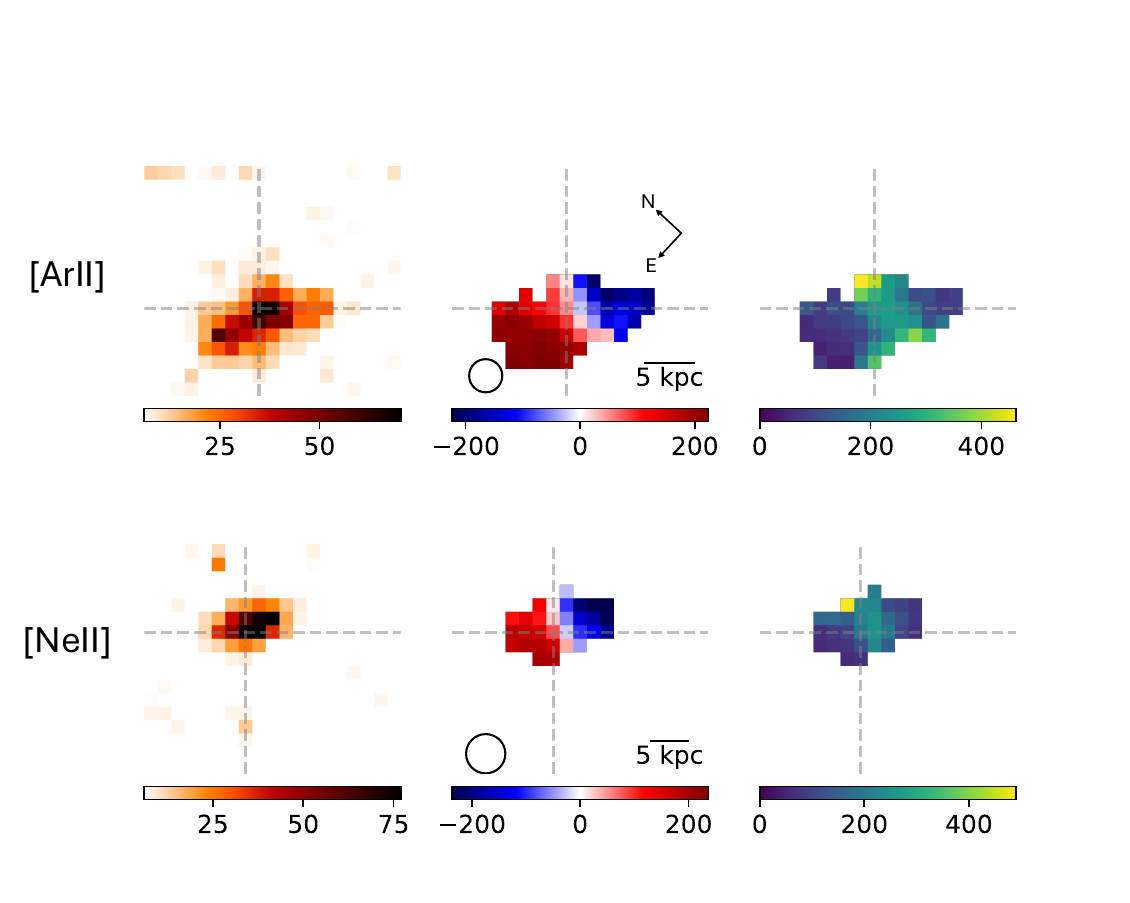}
\caption{Comparison of the kinematic maps from [Ar II] (top), S(5) (middle) and [Ne II] (bottom) for FLS 4.
The columns show the S/N ratio, velocity, and velocity dispersion maps for each line. Gray dashed lines indicate the galaxy center, defined as the peak of the [Ar II]6.99$\mu$m flux.
\label{fig:f4_kmaps}}
\end{minipage}
\end{figure}



\section{Effects of Beam-smearing}
\begin{figure}
\centering
\begin{minipage}[t]{\textwidth}
\centering
\includegraphics[scale=0.6]{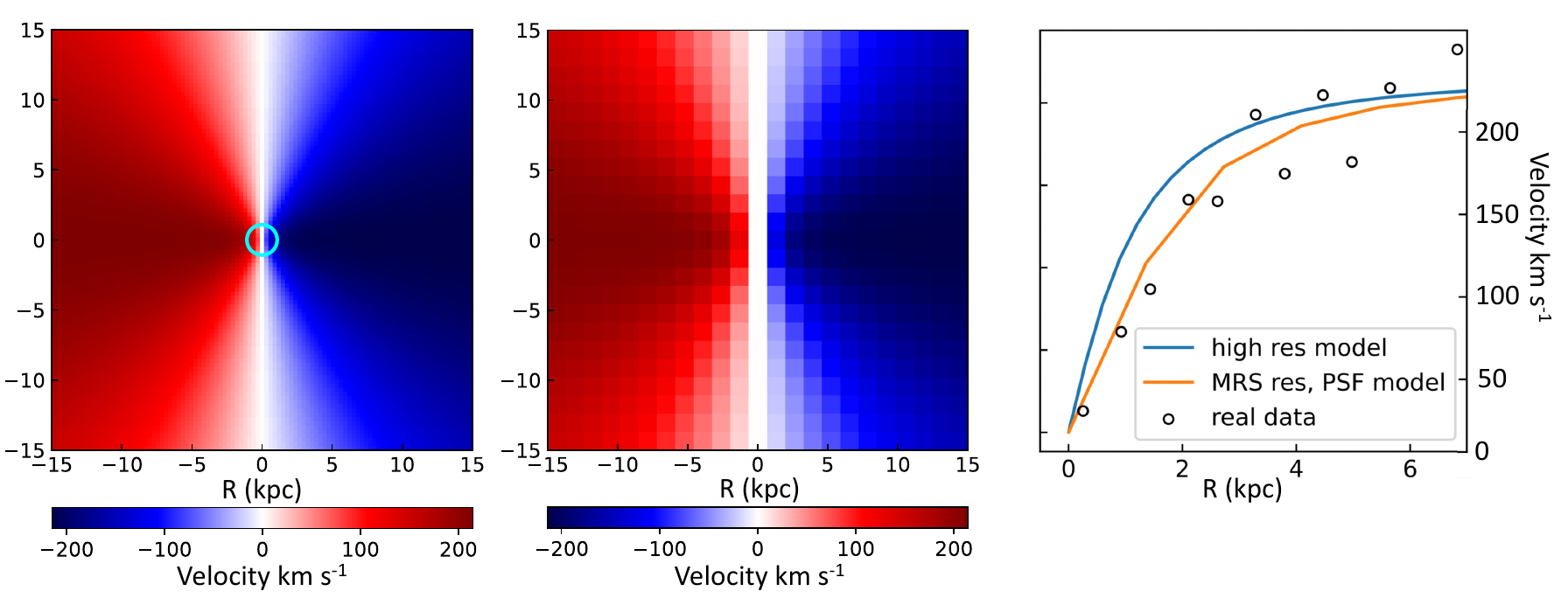}
\caption{Left: High-resolution model velocity field for FLS 4. The PSF-sized aperture for [Ar II] is indicated by the cyan circle at the center of the field.
Center: The same model velocity field, resampled to match the pixel scale of MIRI/MRS and convolved with the MIRI/MRS PSF. Right: Rotation curves extracted from the MIRI/MRS-resolution model (orange), the high-resolution model (blue), and the observed data (black circles).
\label{fig:model}}
\end{minipage}
\end{figure}

One factor that can complicate the interpretation of elevated central velocity dispersion is the beam-smearing of the velocity gradient in the inner regions of these galaxies. This effect arises from the finite angular resolution, which artificially broadens the measured velocity dispersion in regions where the true velocity gradient is steep. To assess the impact of this effect, we constructed a high-resolution model of the line-of-sight velocity field ($V_{los}$), using the equation:
\begin{equation}
    V_{los} = V_{rot}sin(i)cos(\theta),
\end{equation}
where $i$ is i the best fit inclination (see Table \ref{vflat_tab}, column 3) and $\theta$ is the azimuthal angle. $V_{rot}$ is defined as:
\begin{equation}
    V_{rot} = v_{flat}\frac{R}{\sqrt{R^2+R_t^2}},
\end{equation}
where $R_t$ is the turnover radius, which ranges from 1--3.5 kpc for these sources as measured from the velocity curves in Figure \ref{fig:vd_curve}. This is shown in Figure \ref{fig:model} (left). We then degraded the high resolution synthetic data to match MRS pixel size, and convolved this with the MIRI/MRS PSF at [Ar II]6.99$\mu$m (Figure \ref{fig:model}, center). We then extract one-dimensional profiles by averaging a narrow horizontal strip of spaxels, typically $\sim$1 spatial resolution element in width (i.e., a few spaxels), along each column. We calculate the absolute values of the velocities to produce folded velocity curves (Figure \ref{fig:model}, right). This procedure is done to match the analysis of the actual data, which is outlined in \S 4.2.1. As shown in Figure \ref{fig:model} (left), the model accurately reproduces the observed velocity field. We then measured the standard deviation of velocities within a PSF-sized region ($\sim$45 high resolution pixels) at the model galaxy centers, shown as a cyan circle in Figure \ref{fig:model} (right). Comparing this beam-smeared estimate to the observed central velocity dispersions, we find that the observed values are 2.5–7 times larger than the model-based estimates. This indicates that beam-smearing alone cannot account for the high central dispersions. An exception is FLS 1, where the observed sigma is only $\sim$ 1.25 times the beam-smearing estimate, suggesting potential contamination. However, we do not interpret FLS 1 as having elevated dispersion towards the center.

\end{document}